\documentclass{aa}  
\usepackage{graphicx}
\usepackage{txfonts}
\usepackage{caption}
\usepackage{subcaption}
\usepackage[dvipsnames]{xcolor}
\usepackage[colorlinks=true, allcolors=blue]{hyperref}
\usepackage{float}
\usepackage{natbib}
\usepackage[titletoc,title]{appendix}

\begin{document}
   \titlerunning{molecular outflows at cosmic noon}
   \title{A search for cold molecular outflows in cosmic noon galaxies}

   \authorrunning{I. Langan et al.}
   \author{I. Langan
          \inst{1,2,3},
          G. Popping
          \inst{2},
          M. Ginolfi
          \inst{4,5},
          S. Weng
          \inst{6},
          F. Valentino
          \inst{2,9,10},
          G. C. Jones
          \inst{9,11},
          J. Scholtz
          \inst{9,10}
          }

   \institute{Centro de Astrobiología (CAB), CSIC-INTA, Carretera de Ajalvir km 4, Torrejón de Ardoz, 28850, Madrid, Spain\\
              \email{imyoko@cab.inta-csic.es}
         \and
             European Southern Observatory, Karl-Schwarzschild-Str. 2, D-85748, Garching, Germany
         \and
             Univ Lyon, Univ Lyon1, Ens de Lyon, CNRS, Centre de Recherche Astrophysique de Lyon (CRAL) UMR5574, F-69230 Saint-Genis-Laval, France
         \and
             Dipartimento di Fisica e Astronomia, Universit\`a di Firenze, Via G. Sansone 1, 50019, Sesto Fiorentino (Firenze), Italy
         \and
             INAF - Osservatorio Astrofisico di Arcetri, Largo E. Fermi 5, I-50125, Firenze, Italy
         \and
             Aix Marseille Université, CNRS, LAM (Laboratoire d’Astrophysique de Marseille) UMR 7326, 13388, Marseille, France
         \and
             Cosmic Dawn Center (DAWN), Denmark
        \and
             DTU Space, Technical Univeristy of Denmark, Elektrovej 327, DK-2800 Kgs. Lyngby, Denmark
         \and
             Kavli Institute for Cosmology, University of Cambridge, Madingley Road, Cambridge CB3 0HA, UK
        \and
            Cavendish Laboratory, University of Cambridge, 19 JJ Thomson Avenue, Cambridge CB3 0HE, UK
             }

   \date{Received September 15, 1996; accepted March 16, 1997}

 
  \abstract
  {The flow of baryons in and out of galaxies is the primary driver for galaxy evolution -- inflows bring fresh gas to galaxies that will eventually compress into molecular gas to form stars and outflows entrain processed gas outside of galaxies. In addition to depleting the gas reservoir of galaxies, outflows also enrich their circumgalactic medium (CGM) -- which can further impact the next stages of gas accretion, resulting in the presence of molecular gas beyond the stellar component of galaxies, out to CGM scales.  
   In this work, we aim to search for cold molecular gas in the CGM of typical main-sequence (MS) star-forming galaxies (SFGs) at cosmic noon ($z_{med}\sim1.3$), where we expect outflows to be particularly prominent.
  Using Band 3 CO(2--1) observations from the Atacama Large Millimeter and submillimeter Array (ALMA), we study the spatial extent of the cold molecular gas of a sample of 26 SFGs, via stacking techniques. We compare this extent to that of the stacked stellar emission of our sample traced by UltraVISTA Ks band observations. We also search for broad wings in the stacked spectrum which can be indicative of ongoing outflows.
  Within the noise level of the observations, we find that the total intrinsic cold molecular gas of our sample spatially extends to scales of $R_{\text{CO}}\sim12$ kpc, similarly to the stellar emission ($R_{\text{Ks}}\sim13$ kpc). We do not find broad wings in the stacked spectrum that could hint at ongoing molecular outflows, but we find a tentative minor excess of CO(2–-1) emission at negative velocities that might be indicative of outflows, where the redshifted gas is optically thick.
   The absence of high-velocity molecular gas suggests that molecular outflows traced by CO(2--1) emission are weak in MS SFGs at cosmic noon. These weak outflows thus fail to expel a significant amount of molecular gas to CGM scales, as indicated by the absence of molecular emission extending beyond the stellar emission region. This lack of CO emission at large radii could also imply that the molecular gas does not survive at such distances.}

   \maketitle
%

\section{Introduction}
The molecular gas reservoir of galaxies gives us insights on key processes driving galaxy evolution, such as gas flows. The molecular gas of star-forming galaxies (SFGs) is a crucial component impacting the evolution of galaxies, as it hosts and provides the primary fuel for star formation. Molecular gas is therefore directly linked to the star formation rate of galaxies, which drives the overall evolution of galaxies. That reservoir is replenished through gas accretion originating from outside galaxies, i.e., on cosmological scales, as well as recycled gas from gas fountains, and depleted via star-formation and reprocessed gas outflowing to the outside of galaxies, due to stellar- and AGN- feedback (e.g., \citealt{Keres2005}, \citealt{Hopkins2012}, \citealt{Somerville2015}). The molecular gas of galaxies thus provides us clues about the conditions and processes that shaped the evolution of galaxies. However, our understanding of the relation between the molecular gas reservoir of galaxies and their outflows is still incomplete.

In the local Universe, observations of molecular outflows show that metal-enriched cold material can be brought to large distances, out to circumgalactic scales larger than $\sim12$ kpc (e.g., \citealt{Alatalo2011}, \citealt{Bolatto2013a}, \citealt{Veilleux2017}, \citealt{Lutz2020}, \citealt{Girdhar2024}). At cosmic noon ($z=1-3$), the situation is less clear as it becomes increasingly difficult to observe gas flows due to their faint nature. Recent studies using carbon monoxide (CO) observations (tracing the molecular gas) find extended molecular gas reservoir on physical scales of $\sim10-100$ kpc at $z\sim2$, (e.g., \citealt{Ivison2011}, \citealt{Emonts2018}, \citealt{Cicone2021}, \citealt{Scholtz2023}, \citealt{Jones2023}, \citealt{Emonts2024}). However, these studies focus on extreme objects such as protoclusters, quasars, AGN-host galaxies, the brightest submillimetre galaxies (SMGs) or galaxy mergers. The interpretation of such extended molecular gas reservoirs is thus not straightforward, as processes other than stellar-driven outflows e.g., mergers, cosmic inflows, interactions may play a role in producing the observed extended molecular gas. Whether stellar-driven outflows are capable of enriching the surroundings of typical high-redshift SFGs with cold enriched gas is still an open question. Focusing on main-sequence (MS) SFGs is highly relevant for our global understanding of galaxy evolution because they represent the population of galaxies responsible for the bulk of the star formation density of our Universe (e.g., \citealt{Speagle2014} and references therein). Therefore, understanding the fate of the cold gas in and out of MS SFGs gives us critical insights into future stages of galaxy evolution.

Beyond cosmic noon, $z>2$, evidence of $\sim10$ kpc scale carbon reservoirs is found around extreme galaxies, such as dusty starbursts, quasars or mergers at various redshifts (e.g., \citealt{DiazSantos2014}, \citealt{Cicone2015}, \citealt{Ginolfi2020}, \citealt{Falgarone2017}, \citealt{Bischetti2019b}, \citealt{Stanley2019}, \citealt{diCesare2024}, \citealt{FuentealbaFuentes2024}), but also around more typical galaxies. Using stacking techniques, \citet{Fujimoto2019} and \citet{Fujimoto2020} first reported the identification of the single ionized carbon atom [CII] 158$\mu$m (hereafter [CII]) emission out to $\sim$10 kpc scales in typical SFGs at $z\sim6$, far beyond the stellar component of those galaxies as traced by the \textit{Hubble} Space Telescope (HST) observations. They interpreted this result as evidence of cold-mode outflows (e.g., \citealt{Murray2011}, \citealt{HeckmanThompson2017}) driving metal-enriched gas to the CGM. Subsequent work by \citet{Fudamoto2022} found the [CII] emission presented in \cite{Fujimoto2019} and \citet{Fujimoto2020} to be more compact, with a beam-deconvolved effective radius $r_e \sim 2$ kpc. Although this more careful analysis revealed the [CII] emission on more compact scales, \citet{Fudamoto2022} still found that emission to be $\sim3 \times$ larger than the stellar component (\citealt{Fujimoto2020}). Other studies followed and confirmed the possibility of extended [CII] emission around typical SFGs at similar redshifts (e.g., \citealt{Pizzati2020}), suggesting the capacity for outflows in typical SFGs to drive metal-enriched gas to their surroundings. But [CII] can originate from multiple gas phases, i.e., ionised, atomic, and molecular (e.g., \citealt{Sargsyan2012}, \citealt{Pineda2013}, \citealt{Rigopoulou2014}, \citealt{Cormier2015}, \citealt{GloverClark2016}, \citealt{DiazSantos2017}, \citealt{Casavecchia2024a}, \citealt{Casavecchia2024b}). Therefore we cannot draw direct conclusions about molecular outflows and molecular gas reservoirs based solely on [CII] observations.

Alternatively, we can use CO observations to trace the molecular gas of SFGs (e.g., \citealt{Carilli2013}, \citealt{Bolatto2013b}). However, CO emission is fainter than [CII], thus more difficult to probe in the outskirts of galaxies where the gas is more diffuse. \citet{Jones2023} and \citet{Scholtz2023} recently showed evidence for extended ($\gtrsim10$ kpc) CO molecular gas around AGN and quasars, indicative of past AGN-driven outflows. This observational effort at high redshifts supports the possibility of observing similar extended metal-enriched gas around MS SFGs at cosmic noon. The existence of molecular outflows around MS SFGs at cosmic noon is thus yet to be confirmed.

While the spatial extent of molecular gas reservoirs can be a sign of previous stages of outflowing gas, due to the observational challenge in detecting such extended reservoirs, observations of broad wings in molecular and atomic emission lines are more commonly used to identify ongoing outflows (e.g., \citealt{ForsterSchreiber2019}, \citealt{Veilleux2020}, \citealt{Weldon2024}). These broad components in spectral lines can be indicative of high-velocity (speeds of hundreds to thousands of km$^{-1}$) gas that is being ejected from the galaxy, likely due to stellar or AGN-driven outflows. These features provide important insights into the kinematics of the gas, offering valuable hints to the presence of feedback-driven gas flows and the potential impact of outflows on the surrounding CGM. While broad wings linked to the presence of galactic outflows have been observed in [CII] emission lines of high-redshift galaxies (e.g., \citealt{Bischetti2019b}, \citealt{Ginolfi2020}, \citealt{Akins2022}), observations of such broad wings in CO emission lines at high redshifts is much less constrained. With the exception of \citet{Barfety2025}, so far, observations of broad CO wings have been limited to the local Universe and to extreme objects (e.g., \citealt{Cicone2012}, \citealt{Combes2013}, \citealt{Fluetsch2019}, \citealt{Veilleux2020} and references therein).


In this work, we search for signs of molecular outflows around MS SFGs at cosmic noon ($z\sim1-2$), where the Universe is at its peak of star formation, thereby, where gas flow activity could also be at its peak. We look for spatially extended molecular gas at large scales ($\gtrsim10$ kpc) and low-brightness high velocity molecular gas in the form of broad spectral wings. We use stacking techniques on a sample of 26 galaxies from \citet{Valentino2020} with CO(2--1) observations, tracing the molecular gas. Finding faint extended CO emission may reveal the presence of a molecular gas reservoir around those galaxies and the contribution by unresolved faint satellite galaxies. Stacking the spectra of our sample can give us clues on the origin of such extended CO emission. The structure of this paper is as follows. In Section \ref{section:observations}, we describe the sample and observations used to conduct this study. Sections \ref{section:analysisspatialextent} and \ref{section:analysiskinematics} detail the different stacking procedures we explored and the results. In Section \ref{section:discussion}, we discuss our results, and we present a summary of this study in section \ref{subsec:summaryandoutlook_stacking}. The cosmology assumed throughout this work follows the $\Lambda$CDM standard cosmological parameters: $H_0 = 70 \text{km}\, \text{s}^{-1}\, \text{Mpc}^{-1}$, $\Omega_m = 0.3$, and $\Omega_\Lambda = 0.7$ \citep{Planck2020}.
   
\section{Observations and sample description}
\label{section:observations}
\begin{figure}
    \centering
    \includegraphics[width=\columnwidth]{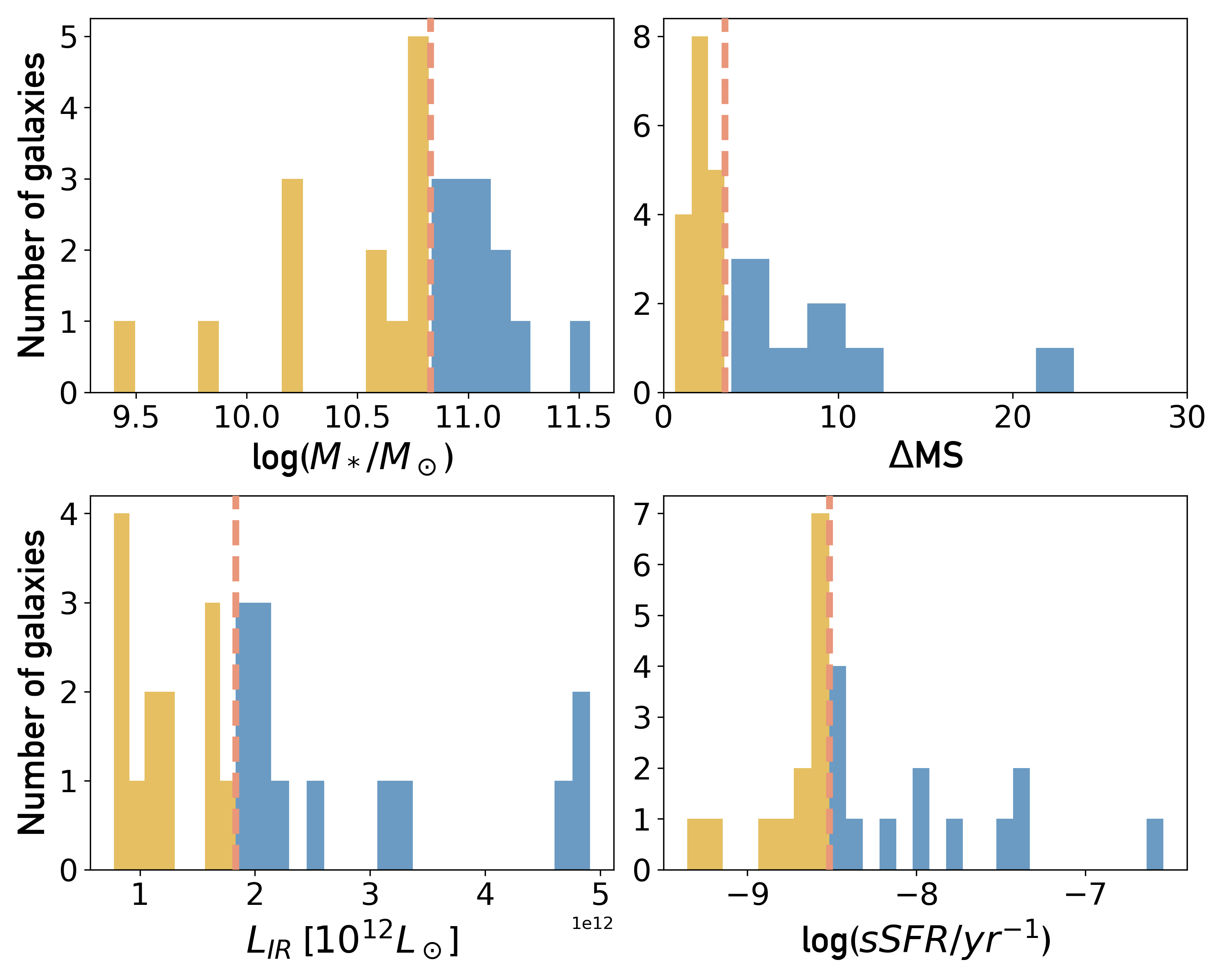}
    \caption{Distributions of the properties of the sample used in this study. Galaxies with values below (above) the median are marked in yellow (blue). The median value is shown with the dashed vertical line, except for the distance to the main sequence for which we show a cut at 3.5 times above (below) the main sequence. Top left: the stellar mass $M_\star$. Top right: the distance to the main sequence as reported in \citet{Valentino2020}. Bottom left: the infrared luminosity $L_{IR}$. Bottom right: the specific star formation rate sSFR. }
    \label{fig:hist}
\end{figure}

We select a sample of galaxies observed with the Atacama Large Millimeter and submillimeter Array (ALMA) in band 3, targeting the CO(2--1) emission line with an average synthesised beamresolution of $1.5\arcsec$ (see \citealt{Valentino2020} for a complete report of the observations). These galaxies are part of an ALMA cycle 4 programme (\#2016.1.00171.S, PI: E. Daddi), designed to observe 75 SFGs (with MS and starburst galaxies) at $z=1.1-1.7$, originally identified by Herschel in the COSMOS area (\citealt{Scoville2007}). Among those 75 SFGs, we select the galaxies with a detection of the CO(2--1) emission line at a signal-to-noise ratio $S/N\geq3$, which results in 33 galaxies. We remove galaxies for which the CO(2--1) emission shows signs of merging (ID 51599, see \citealt{Langan2024}) or for which no stellar mass measurement is reported (IDs 818, 25015) in \citet{Valentino2020} due to a lack of robust constraints in the optical (see \ref{table:technical} for detailed technical information regarding the observations). In the end, our sample contains 26 galaxies, with CO(2--1) emission line detection with $S/N\geq3$, stellar mass and infrared luminosity measurements (see table \ref{table:sample}). In Figure \ref{fig:hist}, we show the distributions of the properties of our sample of typical SFGs. Our sample has the following properties, redshifts $z=1.15-1.63$, stellar masses $M_\star = 10^{9.4-11.5}$, infrared luminosities $L_{IR} = 10^{11.8-12.7} L_\odot$ and star formation rates SFR $ = 115-731 M_\odot$/yr. Table \ref{table:sample} summarises the properties of our sample.

This sample fits the goals of this study for multiple reasons. 1) It provides CO(2--1) emission line observations, one of the lowest CO transitions observable at cosmic noon, allowing us to directly trace the bulk of the molecular gas. Lower CO transitions are more reliable tracers of the bulk of the molecular gas of galaxies because CO transition ratios involving high CO transitions suffer from more uncertain assumptions (e.g., excitation conditions). 2) These observations are taken with a homogenous resolution, $\sim1.5\arcsec$, allowing a straightforward stack without introducing beam corrections prior to the stack. 3) A resolution of $\sim1.5\arcsec$ ($\sim12$ kpc) is small enough to resolve faint emission on extended scales ($\gtrsim 12$ kpc) at the targeted redshifts. 4) The galaxies in this sample are representative of the global population of SFGs, being mostly MS SFGs. 5) The sample is located in the COSMOS field, therefore it is covered by a wide range of multi-wavelength observations, enabling us to put our result in context of galaxy properties, such as stellar mass or infrared luminosity. We specifically use the IRSA COSMOS cutout service\footnote{\url{https://irsa.ipac.caltech.edu/data/COSMOS/index_cutouts.html}} to generate cutouts of the VISTA telescope data in the with VIRCAM/Ks filters (see \citealt{McCracken2012} for details on the ESO programme ID 179.A-2005) of our sample (except for ID 34023 for which the data is missing, see table \ref{table:sample}).

We use the Common Astronomy Software Applications (CASA 6.5.5.21) data processing software \citep{CASA2022} to perform the following data reduction steps, in a homogenised manner on all galaxies of our sample. We subtract the continuum in the uv-plane with a polynomial of order 0 (\texttt{uvcontsub}). We image the CO(2--1) emission with natural weighting and channel width $\Delta v=30 \text{km\,s}^{-1}$ (\texttt{tclean}). This results in cleaned CO(2--1) cubes with an average sensitivity of 0.60 mJy/beam in channels of $30 \text{km\,s}^{-1}$, with an average beam size of 1.5$\arcsec$ ($1.6\arcsec \times 1.3\arcsec$). For 5 galaxies the ALMA observations consisted of multiple executions taken with very different array configurations, resulting in large differences in the beam properties between the individual executions. For those 5 galaxies we only use the observations providing a beam size of $\sim 1.5\arcsec$ to stay consistent with the rest of the sample, which results in cleaned CO(2--1) cubes with an average sensitivity of 0.73 mJy/beam in channels of $30 \text{km\,s}^{-1}$.

\section{Results: the stacked CO intensity map}
\label{section:analysisspatialextent}

\subsection{Intensity map stacking}
\label{subsec:intensity}
We perform the following steps to image the intensity (moment 0) map from each CO(2--1) cube in our sample. 1) We extract the spectrum using a 2$\arcsec$ radius aperture located at the centre of the cube. 2) We randomly perturb the spectrum by its 1 $\sigma$ noise level and fit a Gaussian to the perturbed spectrum. We repeat this noise perturbation procedure 500 times and take the median fit of those 500 iterations as the final best Gaussian fit. 3) We create the intensity map using $\pm$FWHM $\pm$ its error $\sigma_{\text{FWHM}}$ of the best Gaussian fit, using \texttt{immoments}. We show examples of the extracted spectra and their fit in the appendix (Figure \ref{fig:examplefits}).\par
We spatially realign each CO(2--1) intensity map to the peak of the emission before stacking each of them. While outflows are often found perpendicular to the major-axis of galaxies and biconical with an opening angle of $\lesssim 40^{\circ}$ (e.g., \citealt{Bordoloi2011}, \citealt{Rubin2014}, \citealt{Schroetter2016}, \citealt{Guo2023}), here we note that we did not perform further morphological alignments, e.g. position angle (PA) alignment due to a lack of information (see discussion in section \ref{section:discussion}). We perform stacking, such that for N=26, the size of our sample, the intensity map stack is

\begin{equation}
\label{eq:stack}
    I_{stack}(x_i, y_i) = \frac{\sum_{n=1}^{N}I_n(x_i, j_i)w_n}{\sum_{n=1}^{N} w_n}
\end{equation}

where $w_n=1$ in the uniform stack case, and $w_n=1/\sigma_n^2$ in the noise-weighted stack case with $\sigma_n$ the standard deviation of intensity map $n$ masking the detection at the center. The noise-weighted stack showed very similar results due to the homogeneity of the observations, therefore we proceed forward with uniform stacks only. We computed the stacked synthesised beam in the exact same way. This procedure resulted in the uniformly-weighted stack intensity map in Figure \ref{fig:stackedmom0}. The stacked intensity map has a noise level of 0.029 Jy/beam $\text{km\,s}^{-1}$ and a CO(2--1) peak detection with $S/N=27$. In Figure \ref{fig:stackedmom0}, we can already observe faint CO(2--1) emission beyond the beam. However, as noted above, we are agnostic to the orientation and inclination of our sample. This apparent emission extending beyond the beam in Figure \ref{fig:stackedmom0} could be due to morphological misalignment rather than spatially extended faint molecular gas emission, as a result of molecular outflows. Therefore, in Appendix \ref{section:apd} we show uniformly-weighted stack intensity maps resulting from randomly rotating the individual intensity maps prior to stacking. In Figure \ref{fig:restacking} we observe that the shape of the CO(2--1) emission can vary, showing that depending on the rotation, the stacked intensity map shows different degrees of resolved faint emission.

\subsection{Surface brightness radial profiles}
\label{subsec:COradialprofiles}
\begin{figure}
    \centering
    \includegraphics[width=0.8\columnwidth]{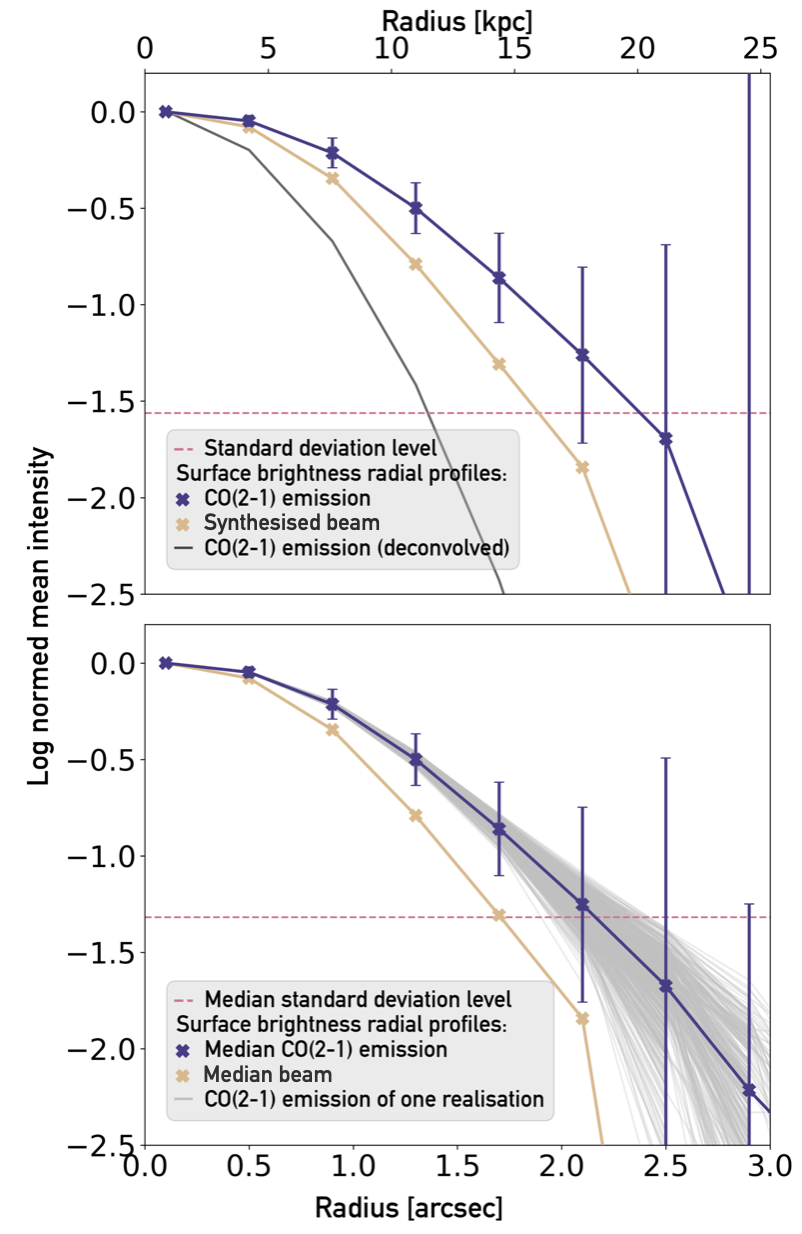}
    \caption{Surface brightness radial profiles of the CO(2--1) emission (in purple) and the synthesised beam (in beige) of the uniform stacked CO(2--1) intensity map. The pink dashed-line shows the noise level of the map. The upper panel shows the profiles of the entire sample of 26 galaxies, and the intrinsic CO(2--1) emission (black line). The bottom panel shows the profiles of 500 resampling iterations by randomly removing 30\% of the sample (grey thin lines) and the median of these iterations (purple line). The error bars show the median of the uncertainty associated with each iteration within each annulus and the horizontal pink dashed-line shows the median standard deviation of the 500 resampled intensity maps.}
    \label{fig:uniformradprofiles}
\end{figure}

We investigate further the CO(2--1) emission, searching for average low-level spatially extended emission, by computing surface brightness radial profiles. While the 2D stacked intensity map might vary depending on the orientation of the galaxies, i.e., their position angle and inclination (Figure \ref{fig:restacking}, left), the circularly average profiles should stay invariant (Figure \ref{fig:restacking}, right), allowing us to probe the extent of the CO(2--1) emission independently of the orientation of the galaxies. To compute these radial profiles, we measure the mean flux normalised by the peak of the emission within annuli of increasing radius $r$ (starting from $r=0$ out to $r=4\arcsec$, i.e., $\sim0-35$kpc at $z_{med}\sim1.3$, the median redshift of our sample) and width of 4 pixels ($\sim0.4\arcsec$) to reduce noise pixel correlation. The associated error is the standard deviation within each annulus. We test the method presented in \cite{Jones2023} (see their Appendix B) to compute the uncertainty in the surface brightness radial profiles, i.e., testing the errors associated with purely Gaussian noise convolved with the beam, but we find that the standard deviation within the annulus gives higher error values. To stay conservative with our uncertainty estimates, we choose the standard deviation within the annulus. The result of the radial profile computation is shown in Figure \ref{fig:uniformradprofiles}, where the surface brightness radial profiles of the stacked CO(2--1) emission and stacked synthesised beam are shown in purple and beige, respectively. We also show the standard deviation level of the stack intensity map with the horizontal pink dashed line. We find that the CO(2--1) surface brightness radial profile is more extended than the one of the synthesised beam, which means there is resolved emission beyond the beam of the observations.\\

To test whether or not only a handful of galaxies drive the observed profiles, we repeat the entire procedure described here 500 times, starting from creating a stacked intensity map but randomly removing 30\% of the sample each time. In the bottom panel of Figure \ref{fig:uniformradprofiles}, we show the result of this resampling. Similarly to the results found for the full sample stack shown in the upper panel of Figure \ref{fig:uniformradprofiles}, all iterations (shown with grey thin lines) are found to be more extended than the synthesised beam. Therefore, we conclude that when performing circular averaging of the observed CO(2--1) emission (ie., the 1D radial profile), we robustly detect resolved emission beyond the beam of the observations. Within the noise limit of the map of the full sample stack (upper panel of Figure \ref{fig:uniformradprofiles}), we find that the CO(2--1) emission extends out to $r_{max}\sim1.9\arcsec$ ($\sim16$ kpc at $z=1.3$), when observational convolution effects are not taken into account.\\

\begin{figure}
    \centering
    \includegraphics[width=0.8\columnwidth]{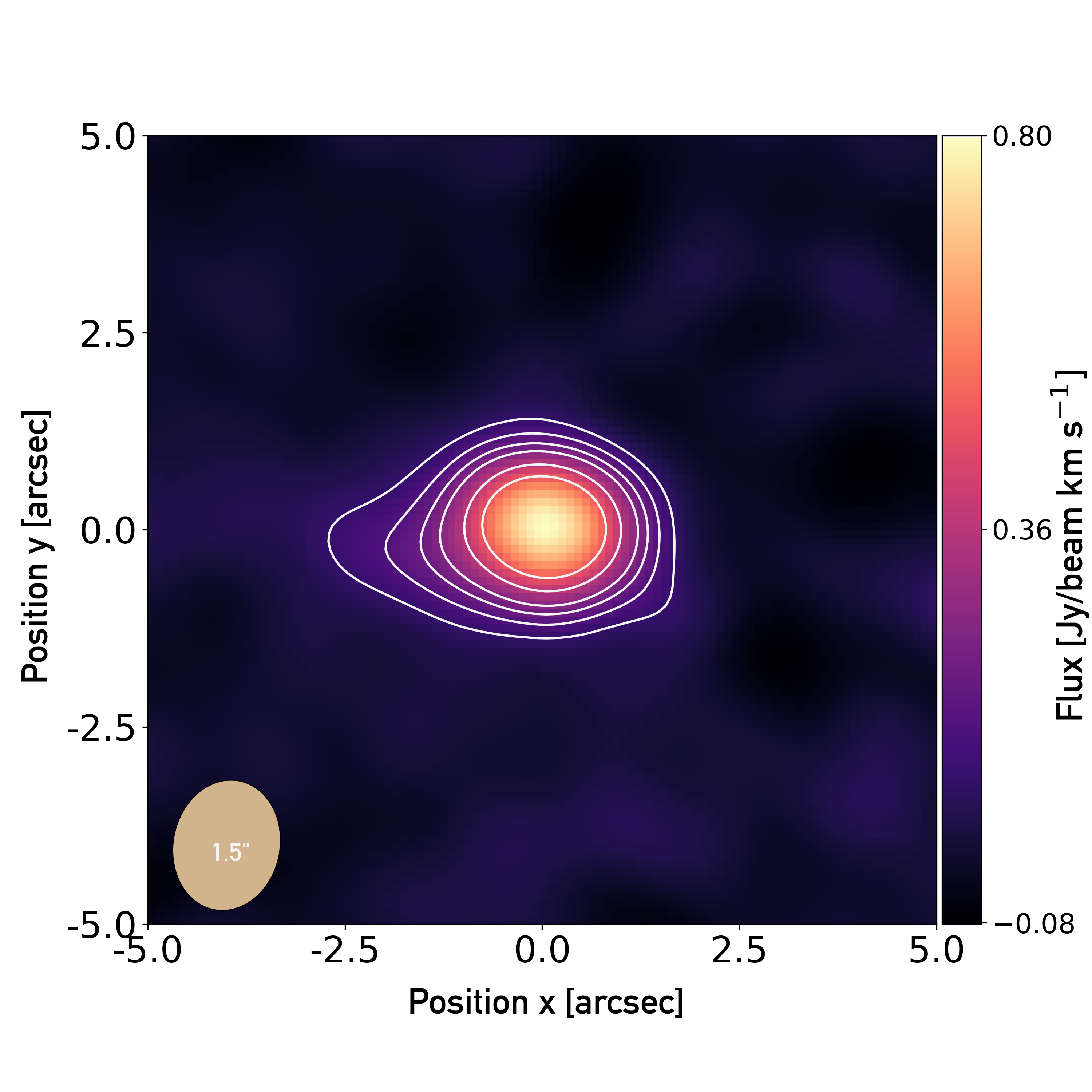}
    \caption{Uniformly weighted stack of CO(2--1) intensity maps. Contours at [3, 5, 7, 9, 13, 17]$\sigma$ are shown with white contours. The beam is shown in the bottom left corner.}
    \label{fig:stackedmom0}
\end{figure}

The observed emission in our stack shown in Figure \ref{fig:stackedmom0} is the result of the intrinsic CO(2--1) emission convolved with the stacked synthesised beam of our observations. To account for the clean synthesised beam convolution introduced during the imaging of the data, and thereby retrieve the intrinsic extent of the stacked CO(2--1) emission, we fit Sérsic profiles (\citealt{Sersic1968}) to the observed CO(2--1) and synthesised beam radial profiles with $R_e$, the half-light radius, $I_e$, the corresponding log normalised intensity and n the Sérisc index, as free parameters. The best Sérsic fits result in $R_{e,\text{CO}}=1.02\arcsec \pm 0.04\arcsec$, $I_{e,\text{CO}}=-0.29 \pm 0.02$ and $n_{\text{CO}}=0.50\pm0.03$ for the CO emission, and $R_{e,\text{beam}}=0.84\arcsec \pm 0.04\arcsec$, $I_{e,\text{beam}}=-0.3\pm 0.03$ and $n_{\text{beam}}=0.50\pm0.04$ for the synthesised beam. Finding $n \sim 0.5$ (i.e., a Gaussian distribution) for both the CO(2--1) and synthesised beam radial profiles, we can apply an analytical 1D Gaussian deconvolution to the observed stacked CO(2--1) emission. In other words, we subtract in quadrature $R_{e,\text{beam}}$ from $R_{e,\text{CO}}$, and find $R_{e,\text{CO}_i}=0.6\arcsec \pm 0.06$, the half-light radius of the beam-corrected CO(2--1) profile. We show the corresponding intrinsic radial profile with the black solid line in the upper panel of Figure \ref{fig:uniformradprofiles}. Within the noise limit of our observations (horizontal pink dashed-line), we find that, on average, the intrinsic CO(2--1) emission extends out to $R_{\text{CO}}\sim1.4\arcsec$ ($\sim12$ kpc at $z_{med}=1.3$), with a half-light radius of $R_e=5$ kpc. 

\subsection{CO cube stacking}
\label{subsec:cubestacking}
\begin{figure}
    \centering
    \begin{subfigure}[b]{0.48\textwidth}
         \centering         \includegraphics[width=\textwidth]{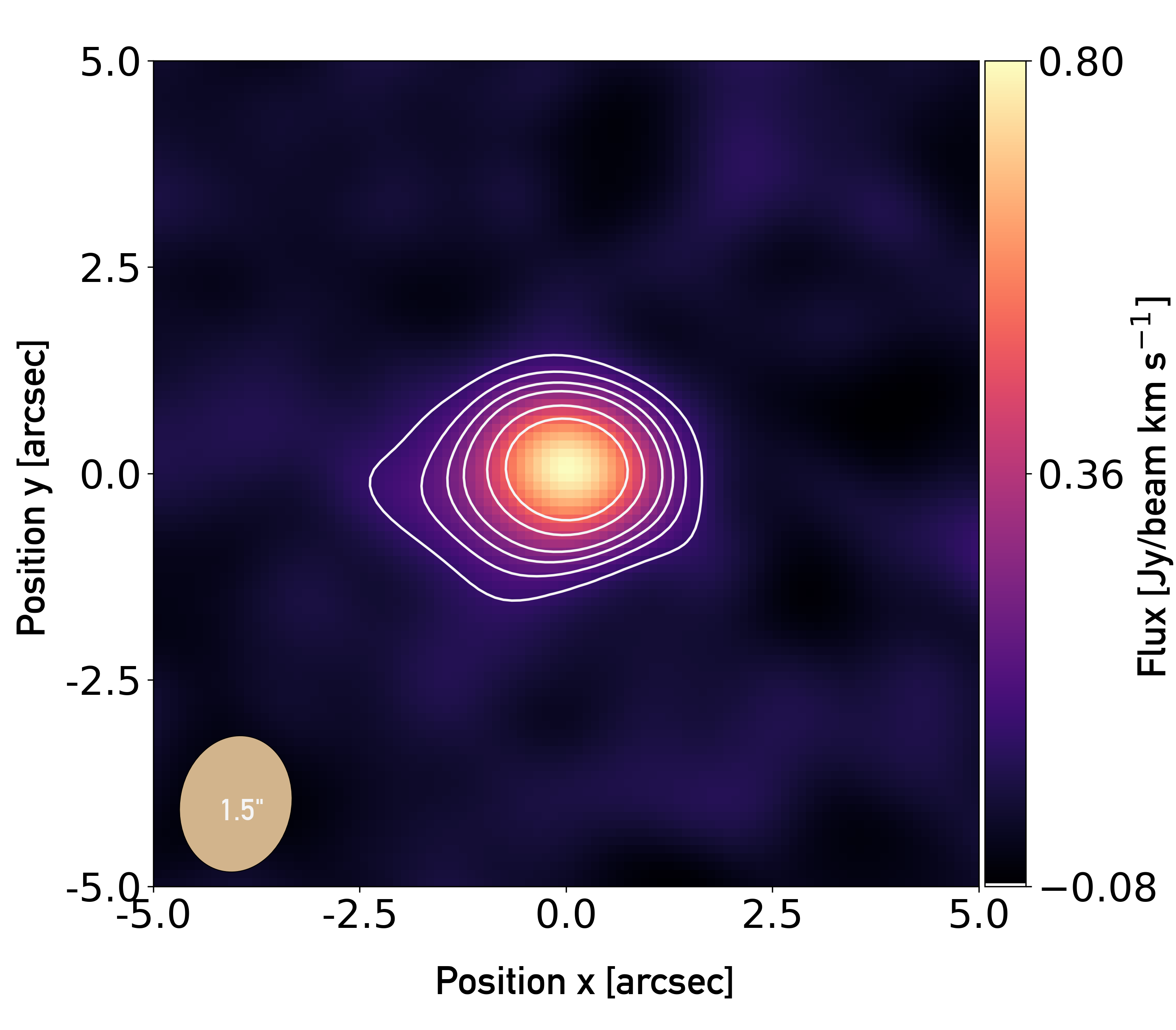}
     \end{subfigure}
     \begin{subfigure}[b]{0.46\textwidth}
         \centering
         \includegraphics[width=\textwidth]{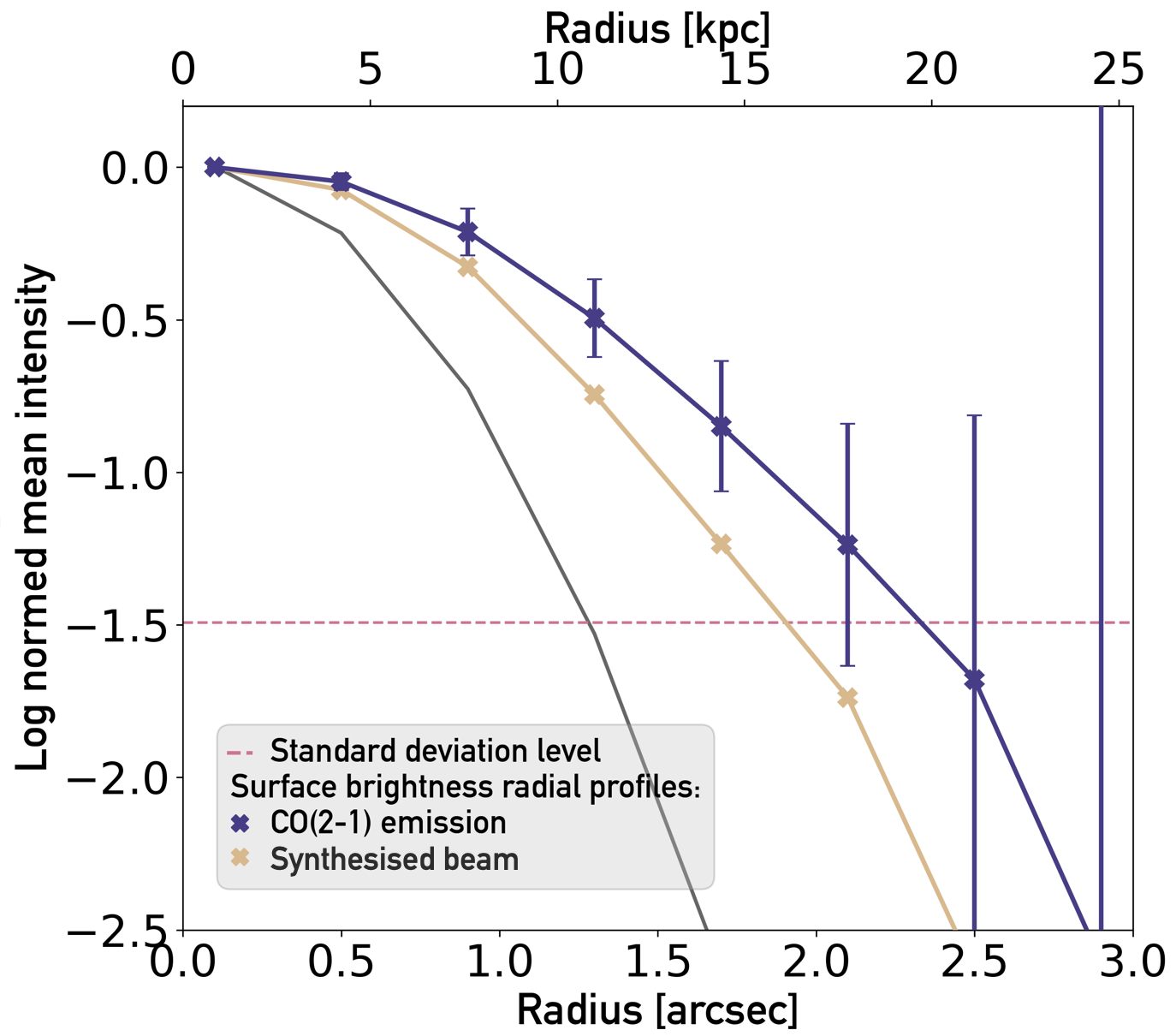}
     \end{subfigure}
\caption{Top panel: CO(2--1) intensity map resulting from the stacked cube (uniformly weighted). The beam is shown in the bottom left corner. Bottom panel: Surface brightness radial profiles of the CO(2--1) emission (in purple), the synthesised beam (in beige) of the CO(2--1) intensity map imaged from the uniformly stacked cubes, and the intrinsic CO(2--1) emission (in black).}
\label{fig:mom0andprofilecube}
\end{figure}

As an alternative to intensity map stacking, we explore cube stacking. We first create a stacked cube from which we then image the CO(2--1) intensity map and derive the corresponding surface brightness radial profiles. For every cube in our sample, we perform the following steps to create the stacked cube. We align each cube in the velocity space to the peak of the CO(2--1) emission, according to the best Gaussian fit as described in step 2) section \ref{subsec:intensity} above. We perform the same velocity alignment to the synthesised beam cube. We align each cube in the position space to the peak of the CO(2--1) emission, according to the peak of that emission in the individual intensity map (created when performing the stack intensity map in section \ref{subsec:intensity}).  We stack the CO(2--1) and corresponding synthesised beam aligned cubes. From the stack CO(2--1) and synthesised beam cubes, we image the CO(2--1) intensity map and compute the corresponding surface brightness radial profile using the same steps as described previously in section \ref{subsec:intensity}. The results are shown in Figure \ref{fig:mom0andprofilecube}. In the bottom panel of Figure \ref{fig:mom0andprofilecube}, we find CO(2--1) emission extending beyond the synthesised beam, out to $\sim1.9\arcsec$ ($\sim16$ kpc) within our noise limit. We perform the synthesised beam correction from the measured CO(2--1) profile (following the same method as described previously) and found the intrinsic CO(2--1) emission extending out to $\sim1.3\arcsec$ ($\sim12$ kpc). Therefore, we observe consistent results between the two intensity maps imaged from stacking the individual intensity maps (Figure \ref{fig:stackedmom0}) or from first stacking the individual cubes (Figure \ref{fig:mom0andprofilecube} top panel).

\subsection{Subsample intensity map stacks}

\begin{figure}
    \centering  \includegraphics[width=\columnwidth]{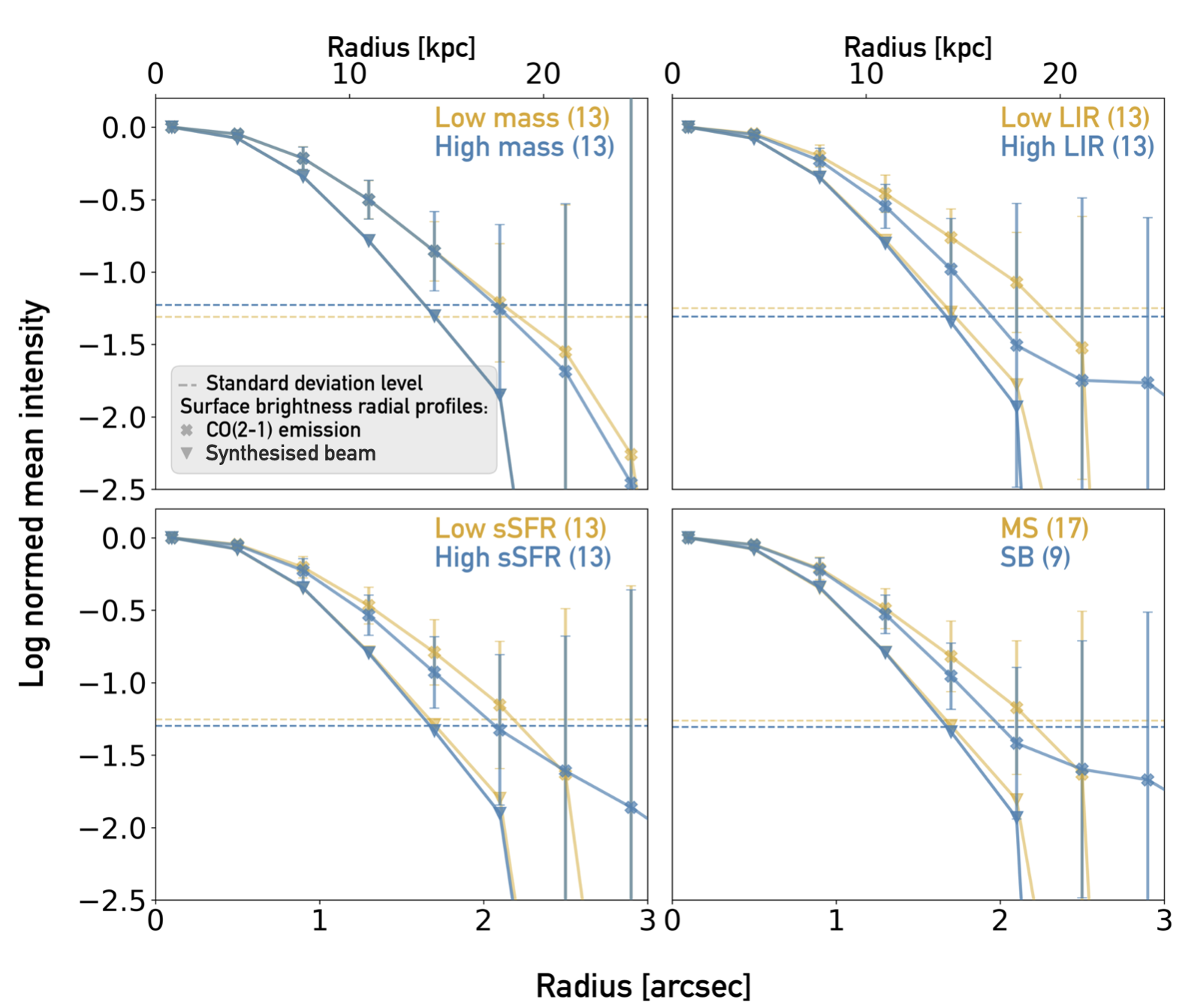}
    \caption{Surface brightness radial profiles of the CO(2--1) emission and corresponding synthesised beam in different galaxy property bins. The number of galaxies in each bin is shown in parentheses. Low (MS) bins are shown in yellow and high (SB) bins are shown in blue. Top left: stellar mass bin, where low mass $\leq10^{10.83} M_\odot$ and high mass $>10^{10.83} M_\odot$. Top right: infrared luminosity bin, where low $L_{IR}\leq10^{12.26} L_\odot$ and high $L_{IR}>10^{12.26} L_\odot$. Bottom left: specific star formation rate bin, where (log) low sSFR $\leq -8.5$ yr$^{-1}$  and (log) high sSFR $> -8.5$ yr$^{-1}$ . Bottom right: distance to the main sequence, where MS $\leq3.5$ times correspond to galaxies within the scatter of the MS and SB $>3.5$ times correspond to galaxies above.}
    \label{fig:radprofilesbins}
\end{figure}

We further investigate potential correlations between the properties of the galaxies in our sample (i.e., their stellar mass ($M_\star$), infrared luminosity ($L_{IR}$), specific star formation rate (sSFR) and distance to the main sequence) and the surface brightness radial profiles. We create binned intensity map stacks according to low (high)-$M_\star$, $L_{IR}$, sSFR, and distance to the main sequence. The "low" and "high" bins are defined with respect to the median value of the sample for the property of interest, i.e., $M_\star = 10^{10.83} M_\odot$ for $M_\star$, $L_{IR} = 1.83\times10^{12} L_\odot$ for $L_{IR}$, and $\log{sSFR} = -8.51$ yr$^{-1}$ for the sSFR (see Figure \ref{fig:hist}). For the distance to the main sequence, we define "MS" (main sequence) and "SB" (starburst) bins according to a distance to the MS of less or more than $3.5$ times, respectively (following the definition set in \citealt{Valentino2020}). We show the stack intensity maps in different galaxy property bins and their CO(2--1) (and corresponding synthesised beam) surface brightness radial profiles in Figure \ref{fig:radprofilesbins}. In all binned stacks, we do not observe a significant difference (i.e., the radial profiles are consistent within their 1 $\sigma$ uncertainty) in the surface brightness radial profile that could favour a galaxy property to drive the observed profiles.

\subsection{Comparison to the stellar component}
\label{subsec:KsStacking}

\begin{figure}
    \centering
    \begin{subfigure}[b]{0.48\textwidth}
         \centering         \includegraphics[width=\textwidth]{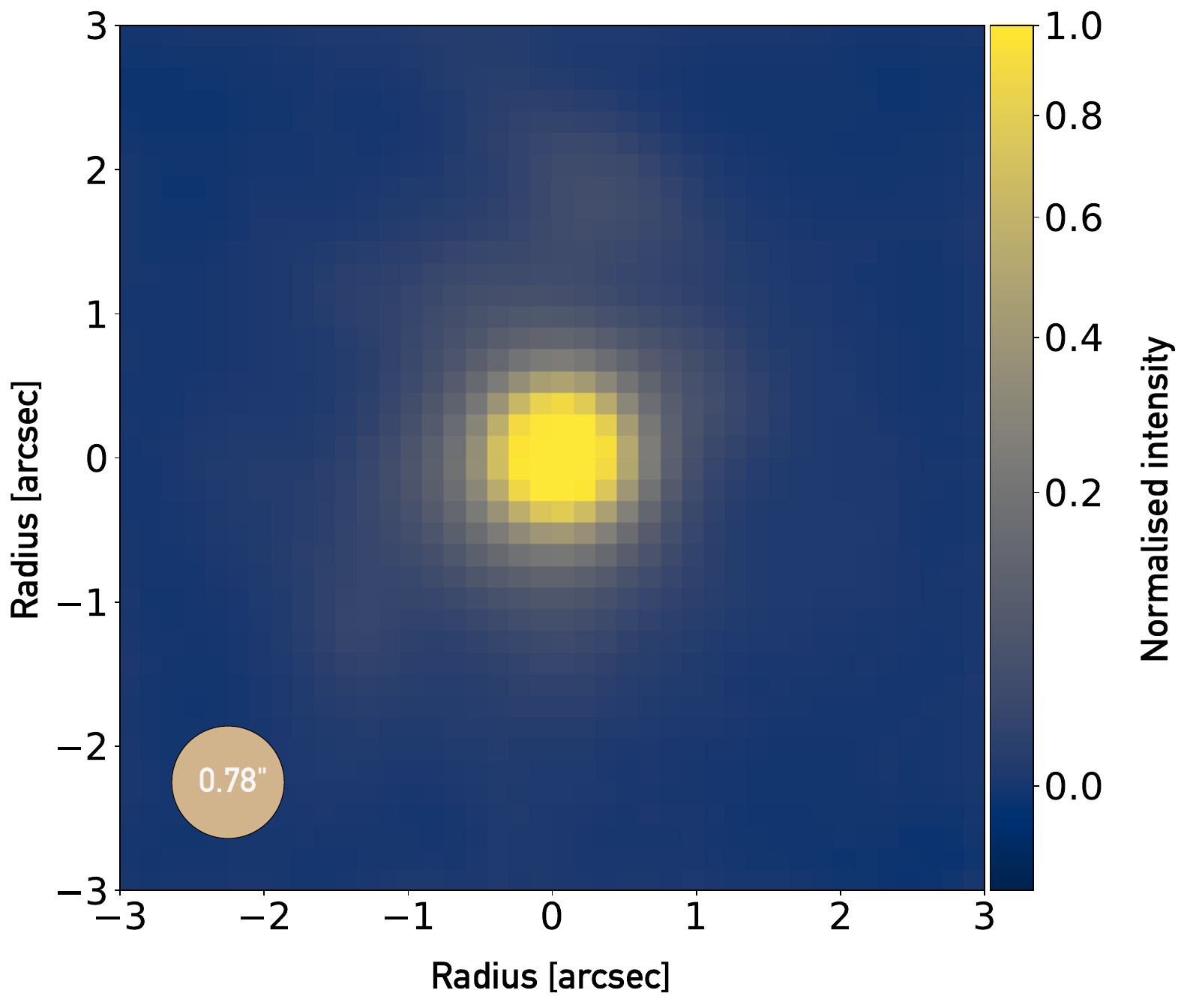}
     \end{subfigure}
     \begin{subfigure}[b]{0.46\textwidth}
         \centering     \includegraphics[width=\textwidth]{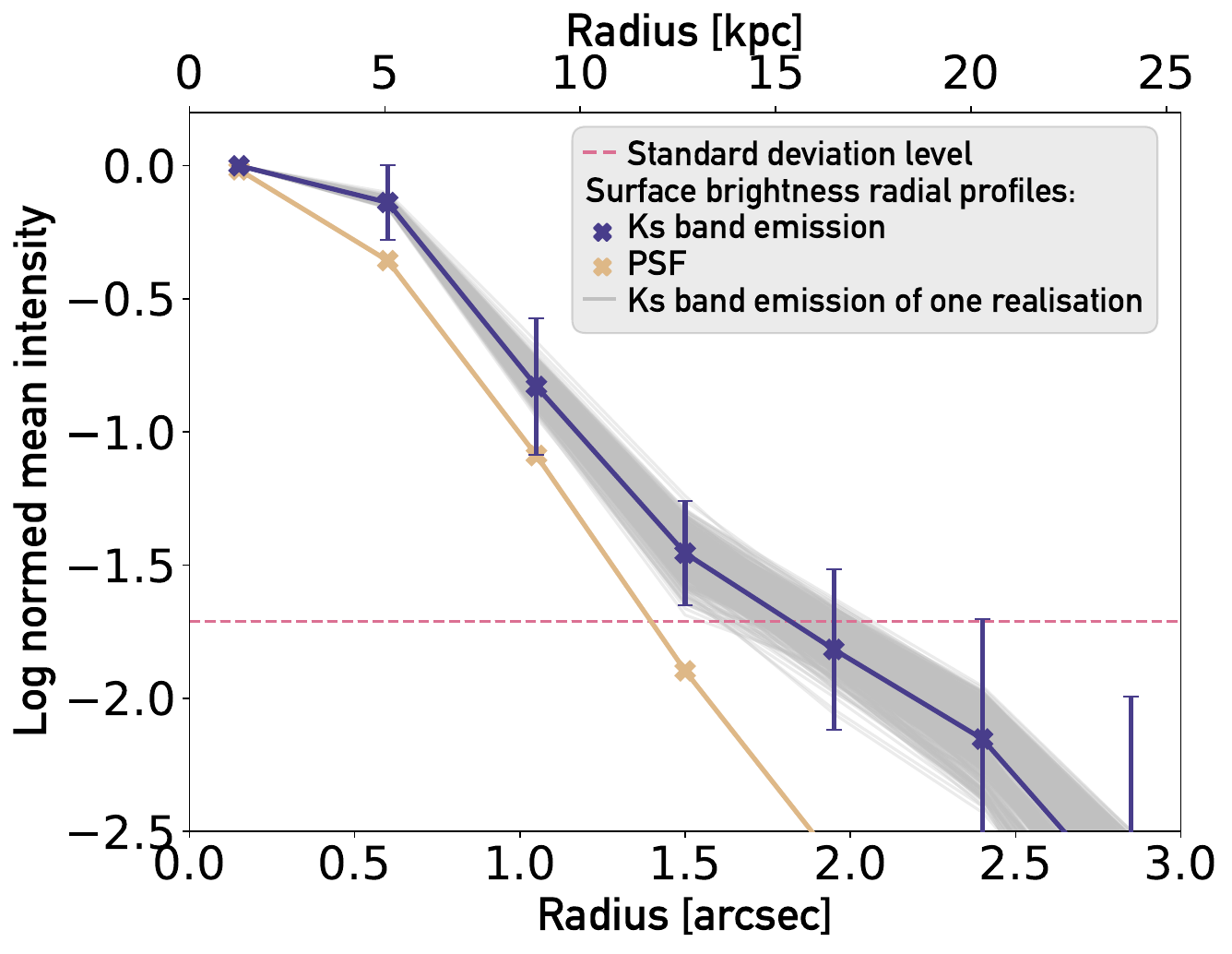}
     \end{subfigure}
\caption{Top panel: Stacked Ks band PSF-deconvolved and background subtracted image. The diameter of the beige circle in the bottom left corner corresponds to the FWHM of the PSF. Bottom panel: Surface brightness radial profiles of the deconvolved stacked Ks band emission (in purple), the corresponding PSF (in beige), and the 3 $\sigma$ noise level of the stacked image (horizontal dashed pink line). The surface brightness radial profile of each deconvolved and resampled stack is shown with grey thin lines.}
\label{fig:KsStack}
\end{figure}

To better understand the physical meaning of the spatial scales at which we find CO(2--1) emission, we compare our CO(2--1) stacked intensity map (section \ref{subsec:intensity}) to the stacked Ks band image of our sample. We realign every Ks band image to its peak emission, i.e., we make a comparison of the extent of the CO(2--1) and Ks band emissions independently. 
We use information provided in \citet{McCracken2012} to compute a 2D PSF Moffat model with FWHM $=0.78\arcsec$ and $\beta=3.5$, and perform a Richard-Lucy PSF deconvolution on each Ks band image using \texttt{scikit-image}. We use \texttt{photutils} (\citealt{Bradley2022}) to create segmentation maps to isolate sources from the background sky, and then we estimate the background noise using the sky pixels only, which we subtract from each image. In two cases, we also use these segmentation maps to mask nearby bright galaxies that do not appear to be located at the same redshift as the central galaxy (based on visually cross-matching with the ALMA CO(2--1) intensity maps). We uniformly stack the realigned, PSF-deconvolved and background-subtracted images and show the result in the upper panel of Figure \ref{fig:KsStack}.

We compute the corresponding surface brightness radial profile, using the same method as for the CO surface brightness radial profile (section \ref{subsec:intensity}). 
In Figure \ref{fig:KsStack}, bottom panel, we show the radial profiles of the stacked deconvolved Ks band emission (purple solid line), the 500 iterations when resampling the stack (thin grey solid lines), the PSF Moffat model (solid beige line) and the 3 $\sigma$ standard deviation level of the stacked deconvolved Ks (dashed pink line). We observe the intrinsic emission in the Ks band, tracing the stellar component of our sample, extending out to $\sim1.5\arcsec$ ($\sim 13$ kpc), down to the noise level of the map.

\begin{figure}
    \centering  \includegraphics[width=\columnwidth]{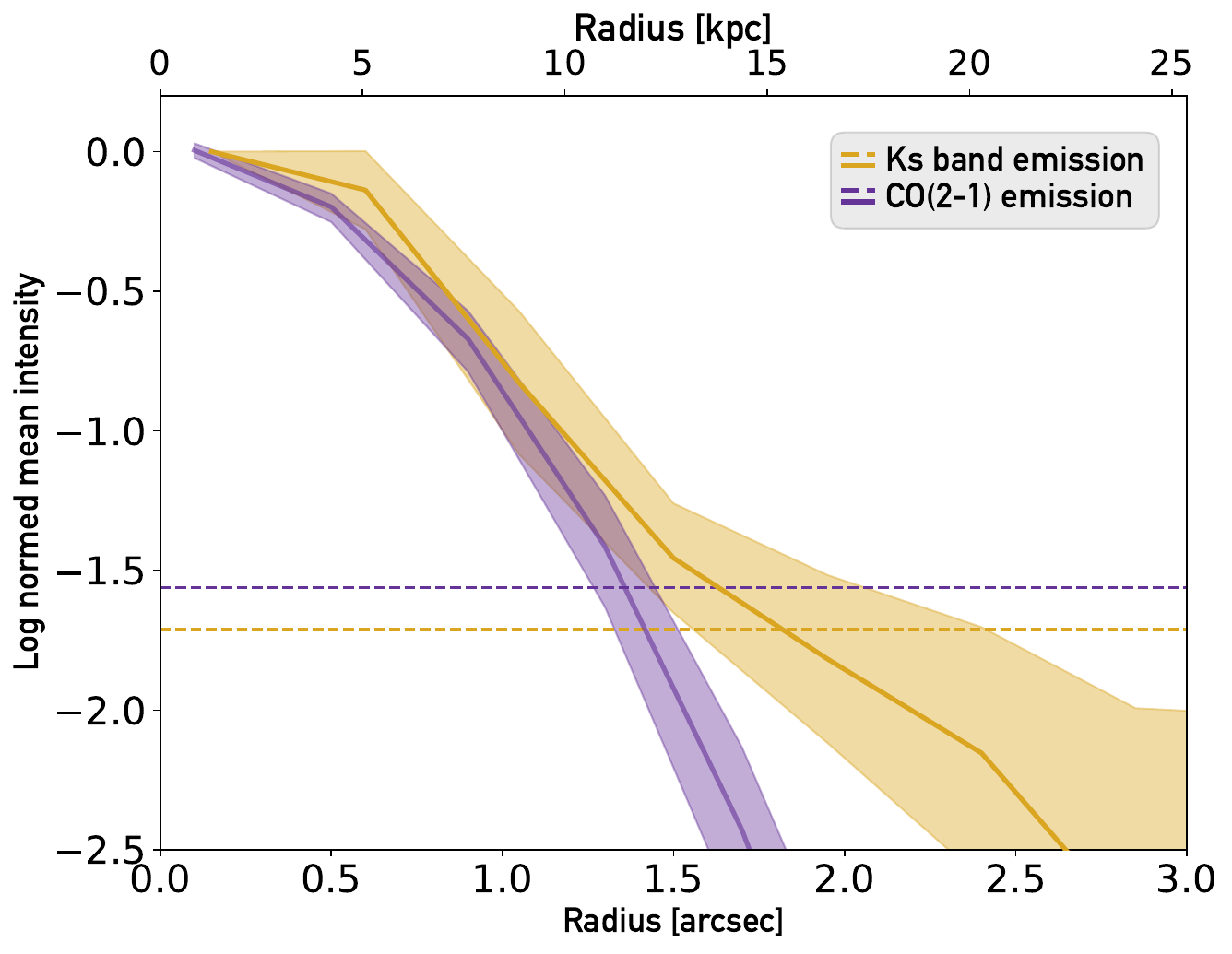}
    \caption{Comparison of the intrinsic radial profiles of the ALMA CO(2--1) (purple) and Ks band (yellow) emissions. The respective standard deviation levels are shown with horizontal dashed line with matching colours. We observe no evidence of extended CO emission compared to the Ks emission, within the noise limits.}
    \label{fig:ALMAKs}
\end{figure}

In Figure \ref{fig:ALMAKs}, we show the comparison of the intrinsic ALMA CO(2--1) emission radial profile (in purple) with that of the Ks band (in yellow), with their respective standard deviation levels (dashed horizontal lines, in purple and yellow). Within the respective noise level of the Ks band and CO stacked maps, we find that the two Ks band and CO radial profiles follow each other closely. The two profiles show a similar extent ($\sim13$ kpc). There is therefore no evidence of CO emission significantly more extended than the Ks band emission.

\section{Results: the stacked CO line profile}
\label{section:analysiskinematics}

\subsection{Spectrum stacking}
\label{subsec:velocity}
\begin{figure}
    \centering
    \begin{subfigure}[b]{\columnwidth}
         \centering
         \includegraphics[width=\columnwidth]{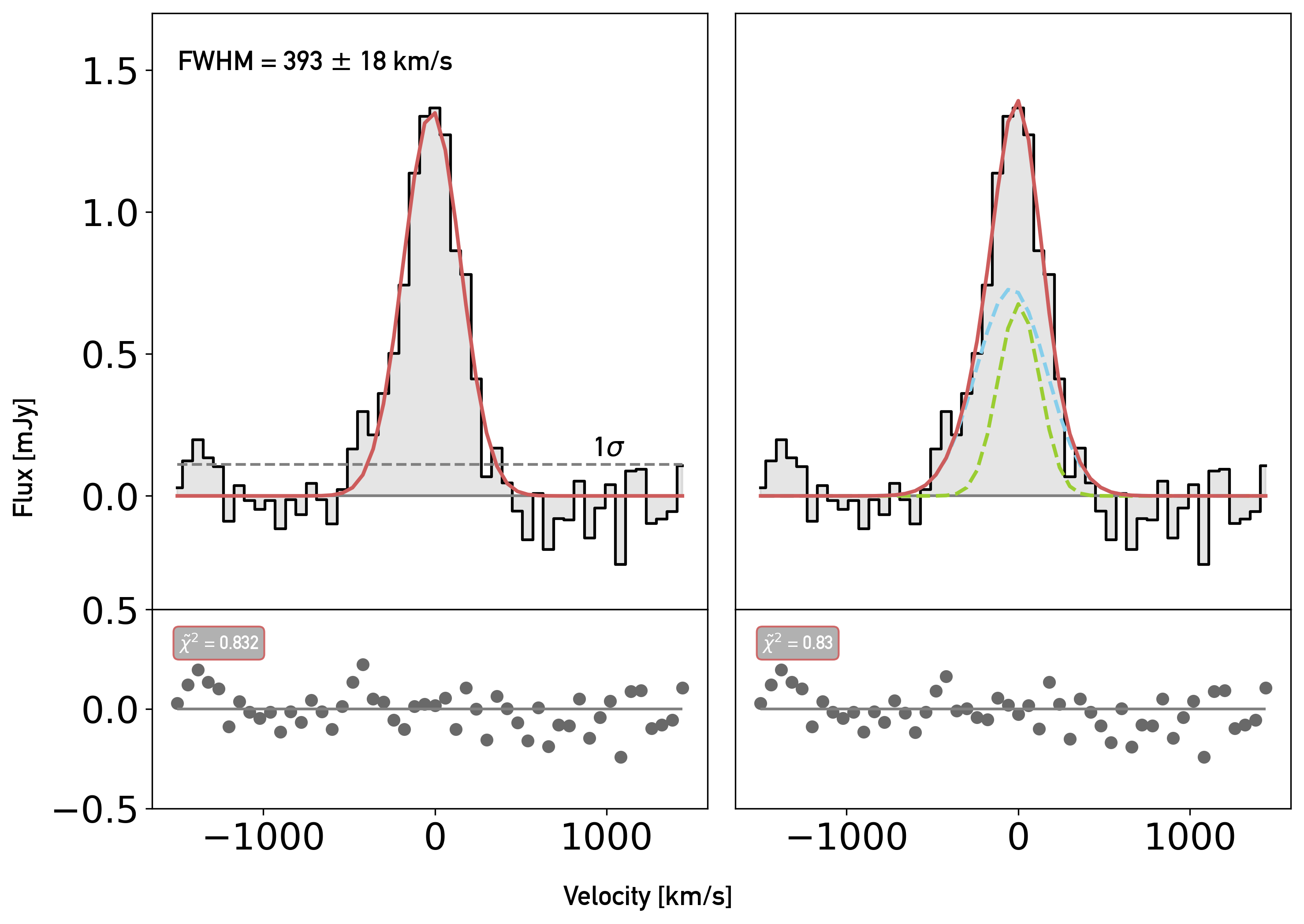}
     \end{subfigure}
     \begin{subfigure}[b]{\columnwidth}
         \centering
         \includegraphics[width=\columnwidth]{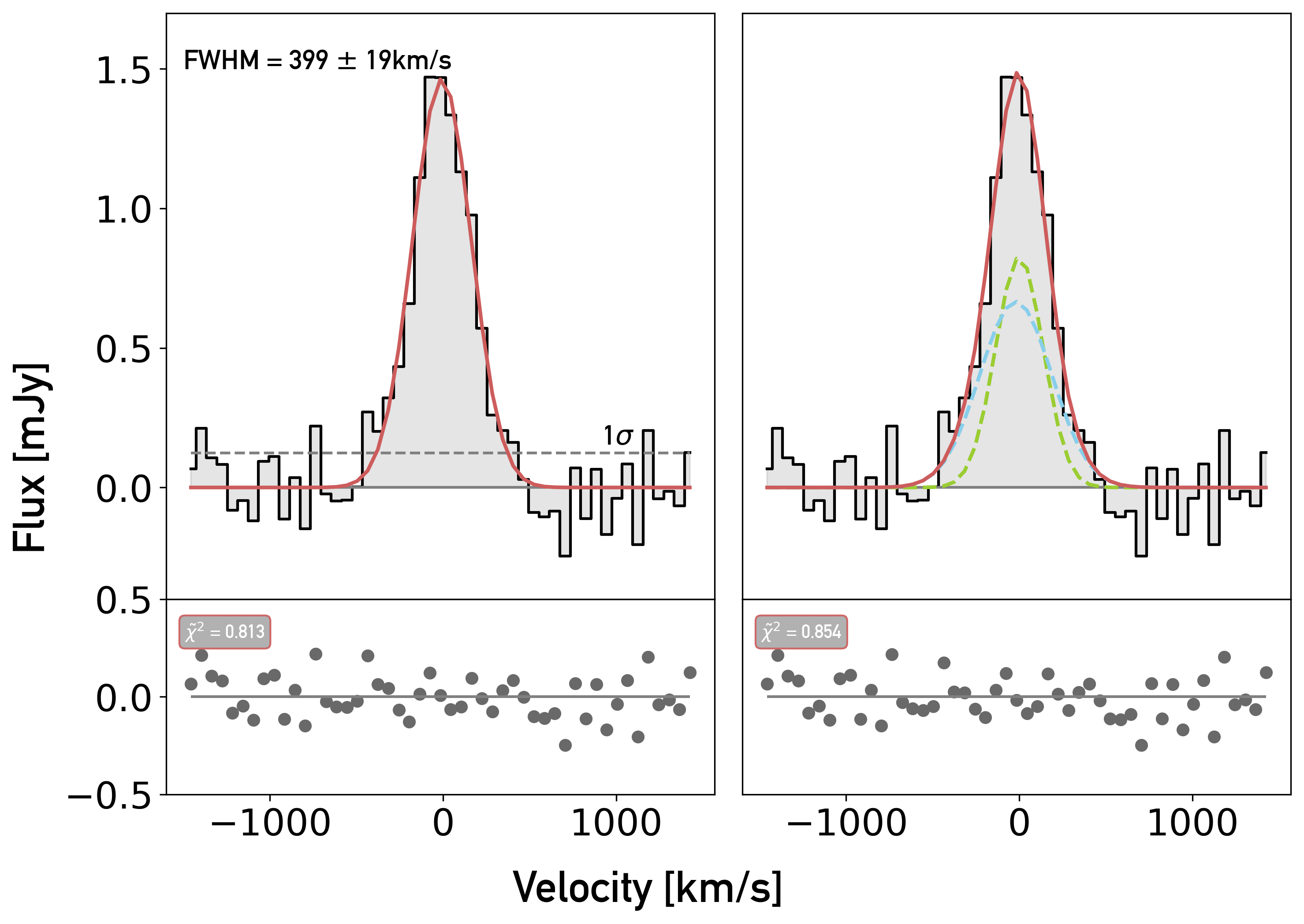}
     \end{subfigure}
\caption{Top panel: Spectrum of the uniformly stacked CO(2--1) emission (in black). Bottom panel: Spectrum of the CO(2--1) emission (in black) extracted from the uniformly stacked CO(2--1) cubes. In the left panel, we show a single-Gaussian fit (in red) and its residuals and corresponding reduced $\chi^2$ in the bottom panel. In the right panel, we show a two-Gaussian fit (the sum of the two Gaussians in red) composed of a narrow component (green dashed line) and a broad component (blue dashed line).}
\label{fig:SpectrumStack}
\end{figure}

We investigate the velocity and velocity dispersion of the CO(2--1) emission line, searching for faint broad wings that could be indicative of outflowing gas. We stack the individual CO(2--1) emission line spectra to gain sensitivity and reveal such potentially faint broad wings. For every cube in our sample, we perform the following steps to create the stacked spectrum. 1) We extract the spectrum using a 2$\arcsec$ aperture centred on the cube. 2) We fit a Gaussian to the resulting CO(2--1) emission line as explained in Section \ref{subsec:intensity}. 3) We use the centre found in the best Gaussian fit to realign each spectrum to the peak of the CO(2--1) emission. 4) We rebin the spectra to 60 $\text{km\,s}^{-1}$ instead of the natural 30 $\text{km\,s}^{-1}$, to reduce noise fluctuations. We perform uniform stacking as in Equation \ref{eq:stack} using the 60 $\text{km\,s}^{-1}$ spectra (all the analysis presented here was also performed on the 30 $\text{km\,s}^{-1}$ spectra, i.e., without rebinning, and is shown in the appendix). The resulting stacked spectrum is shown in Figure \ref{fig:SpectrumStack}, upper subfigure.

We find a stack CO(2--1) emission line with a Gaussian shape, a peak flux of 1.36 mJy, a line width FWHM$=393\pm18$ km$^{-1}$ and a noise level of 0.11 mJy (represented by the horizontal grey dashed-line in Figure \ref{fig:SpectrumStack}, upper subfigure). The average line width and noise level of the individual spectra are $397$ km$^{-1}$ and 0.51 mJy, showing the strength of the stacking method to reduce noise levels. To test the presence of broad wings in the stack spectrum, we perform single-Gaussian (red line, left panel of Figure \ref{fig:SpectrumStack}, upper subfigure) and two-Gaussian fitting (red line, right panel of Figure \ref{fig:SpectrumStack}, upper subfigure, where the narrow and broad components are shown in green and blue dashed-lines, respectively). We compute the residuals and the corresponding reduced chi-squared, $\tilde{\chi}^2$, of each fit and show them in the bottom panels of Figure \ref{fig:SpectrumStack}, upper subfigure. If the combination of a narrow component and a broad component (two-Gaussian fit) results in a fit with a significantly higher quality ($\tilde{\chi}^2$ closer to 1), it is indicative of outflowing material. In the single-Gaussian fit, $\tilde{\chi}^2=0.832$ and in the two-Gaussian fit, $\tilde{\chi}^2=0.830$. Therefore, we find no difference in the quality of the fits. We note the presence of weak flux excess at negative velocities.

To test the systematics of the stacked spectrum, we repeat the uniform stacking of the spectra, 500 times, randomly removing 30\% galaxies of the full sample at each iteration. We repeated the single-Gaussian and two-Gaussian fitting procedure, and found a marginal improvement of the fit in the single-Gaussian case with $\tilde{\chi}^2 = 0.835$ ($\tilde{\chi}^2 = 0.816$ for the two-Gaussian fit). The lack of broad wings in both the full sample stack and resampled stack spectra thus indicate the absence of prominent strong outflowing material. However, we cannot exclude that the S/N of our stacked spectra might prevent us from detecting an outflow signature (see discussion in Section \ref{section:discussion}).


\subsection{CO cube stacking}
We investigate the outcome of stacking individual cubes first, to then create the stacked spectrum. Starting from the stacked cube (see section \ref{subsec:cubestacking}), we perform the same steps described previously in section \ref{subsec:velocity} to create the stacked spectrum. We show the result in the bottom row of Figure \ref{fig:SpectrumStack}. The resulting spectrum show similar results to the one obtained from stacking the individual spectra, i.e., we find a Gaussian-shaped CO(2--1) emission line with a peak flux of 1.46 mJy, line width FWHM $=399 \pm 19$ km$^{-1}$ and noise level of 0.12 mJy. The two-Gaussian fit is marginally better than the single-Gaussian fit, with $\tilde{\chi}^2 = 0.854$ and $\tilde{\chi}^2 = 0.813$, respectively. Therefore, the two methodologies, i.e., stacking individual cubes first, or stacking the individual spectra first, to then create a stacked spectrum, show comparable results.

\subsection{Subsample spectrum stacks}
\begin{figure*}
    \centering
    \begin{subfigure}{\columnwidth}
        \includegraphics[width=\linewidth]{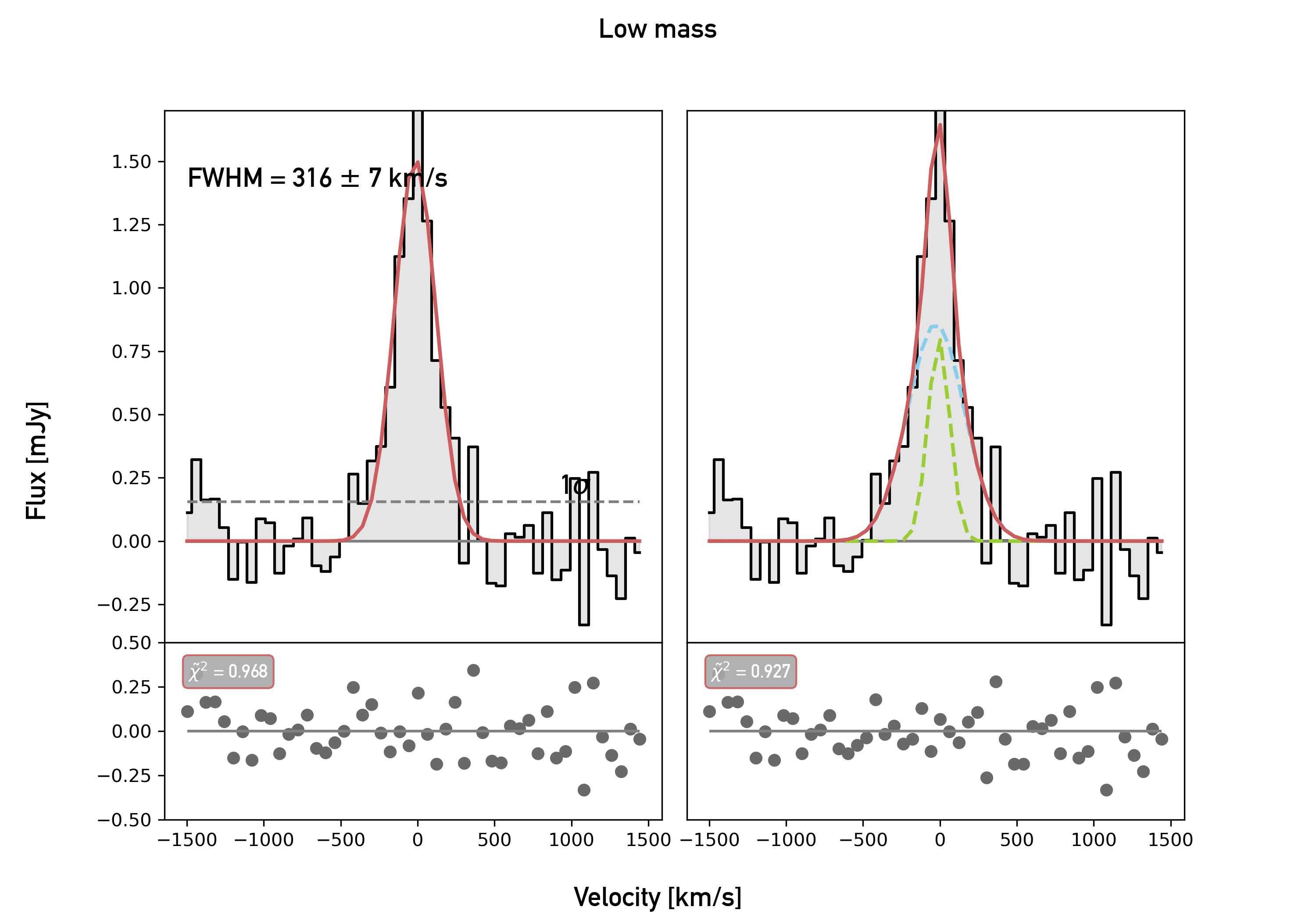}
    \end{subfigure}
    \begin{subfigure}{\columnwidth}
        \includegraphics[width=\linewidth]{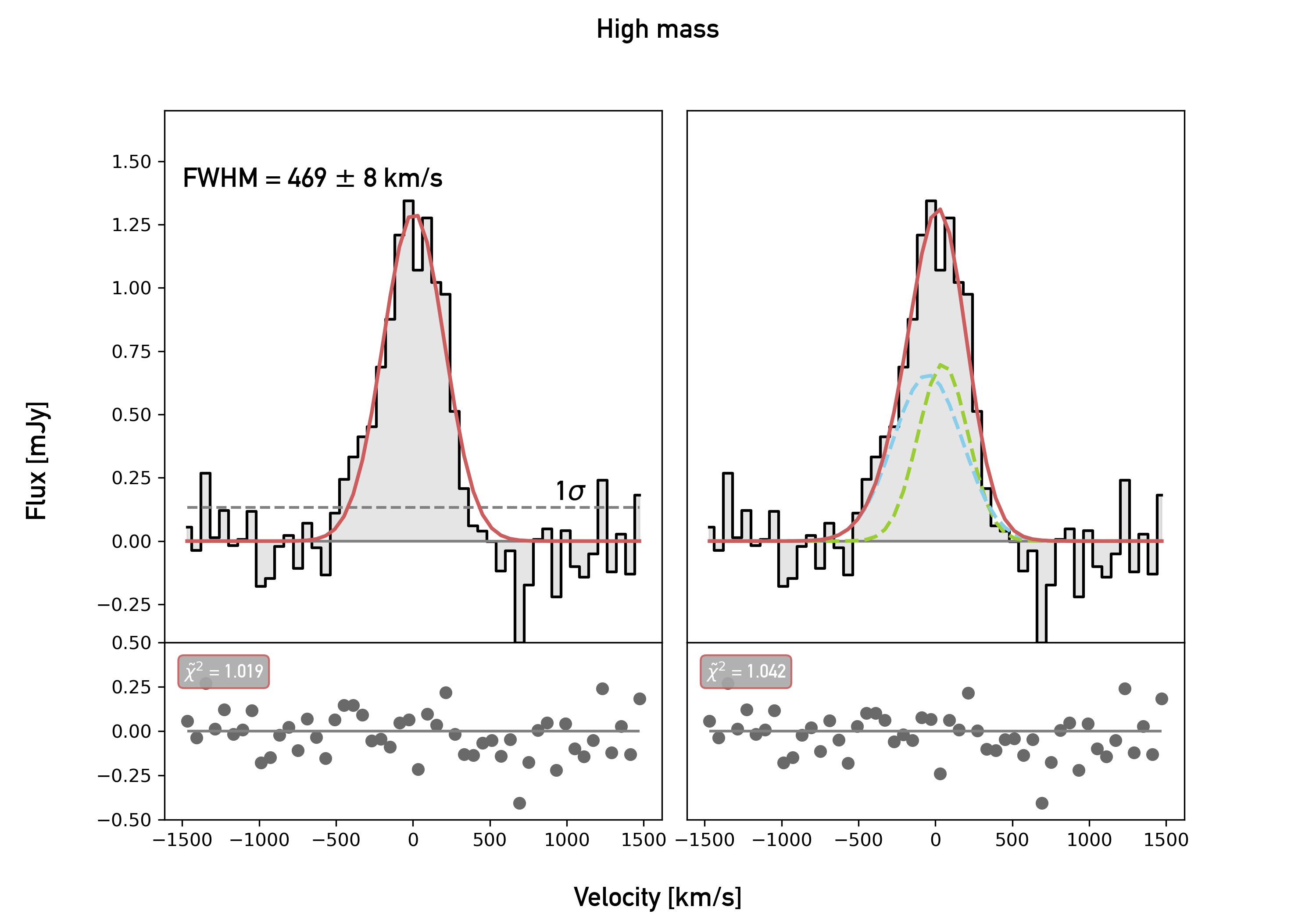}
    \end{subfigure}
    \begin{subfigure}{\columnwidth}
        \includegraphics[width=\linewidth]{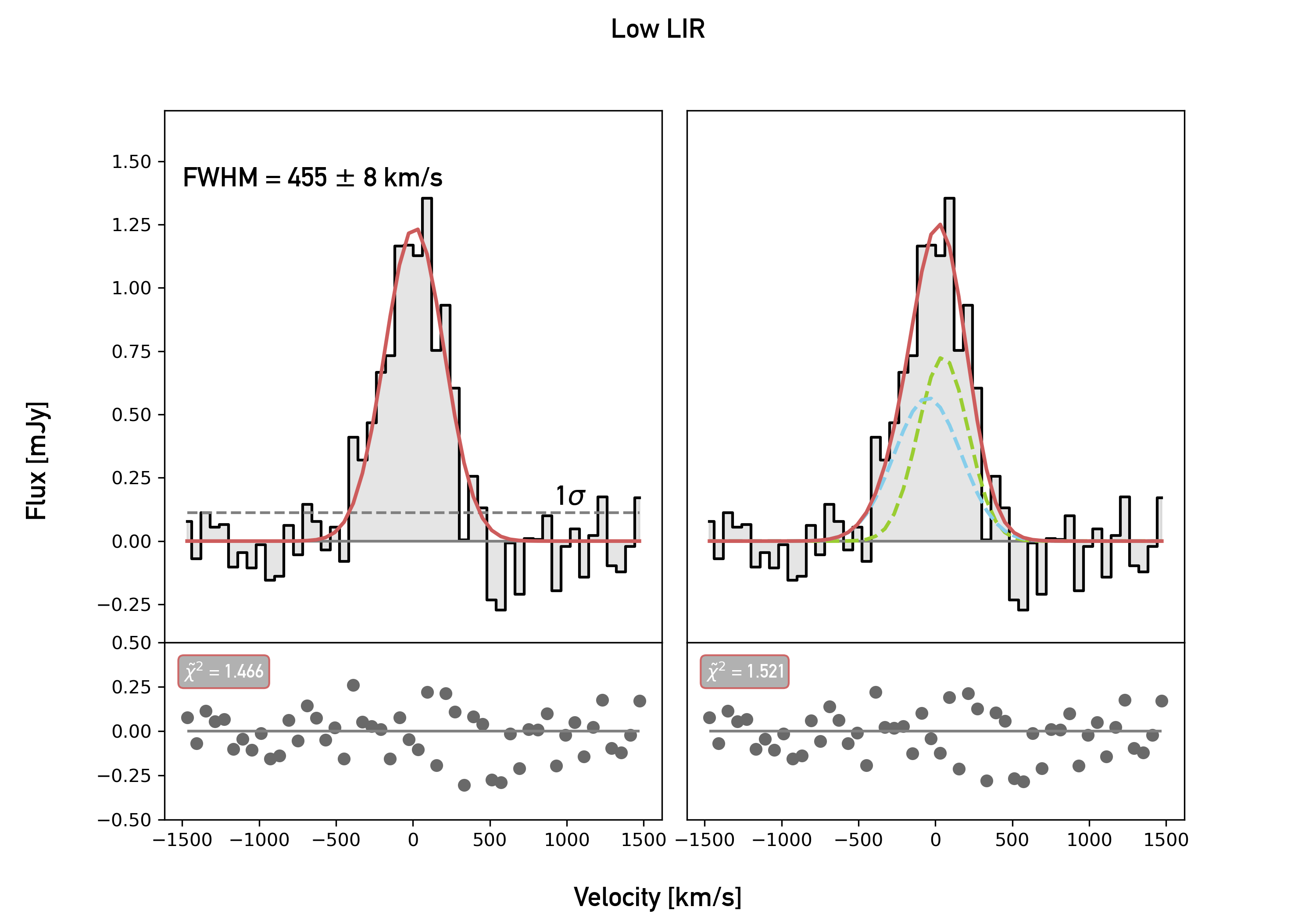}
    \end{subfigure}
    \begin{subfigure}{\columnwidth}
        \includegraphics[width=\linewidth]{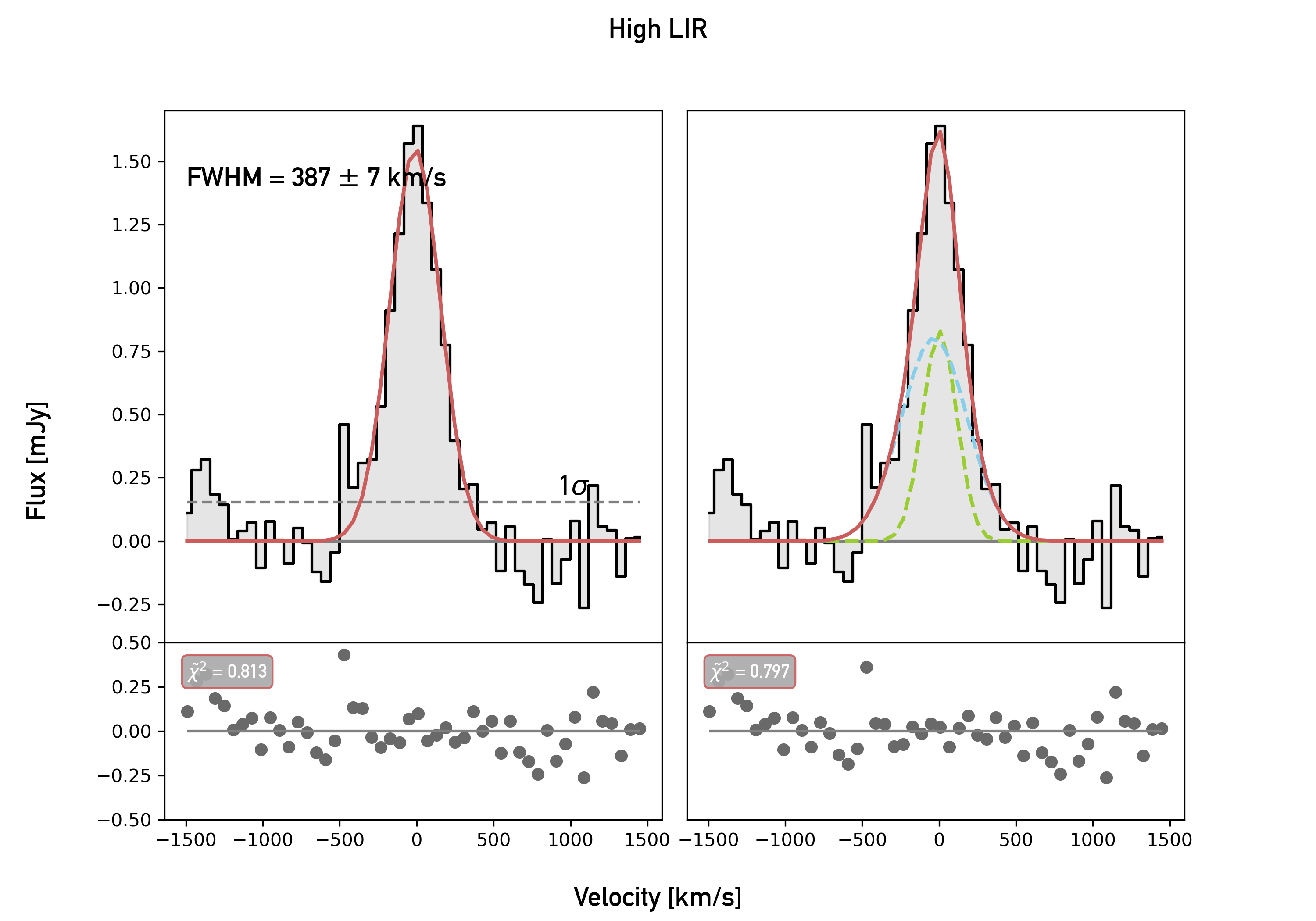}
    \end{subfigure}
    \begin{subfigure}{\columnwidth}
        \includegraphics[width=\linewidth]{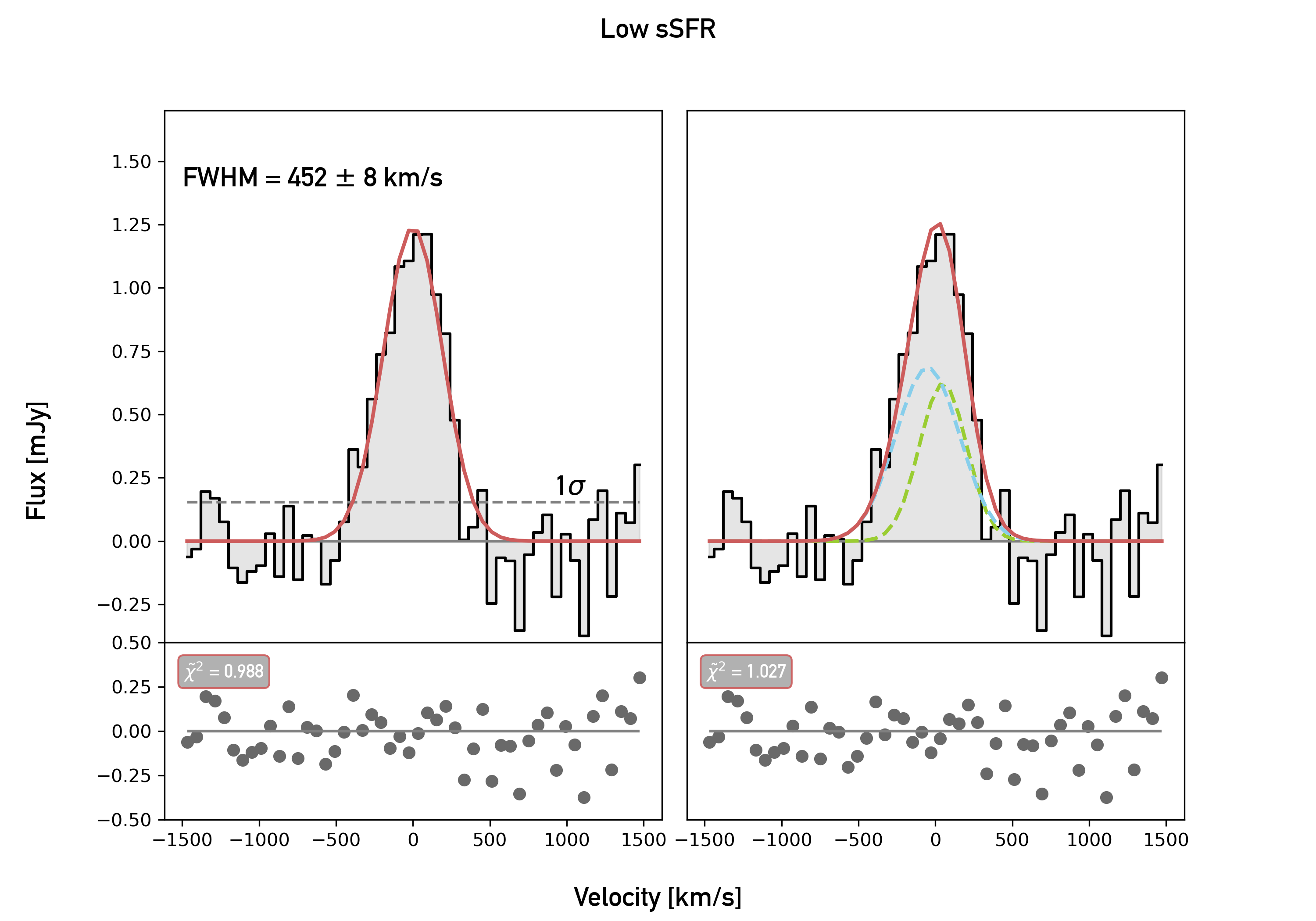}
    \end{subfigure}
    \begin{subfigure}{\columnwidth}
        \includegraphics[width=\linewidth]{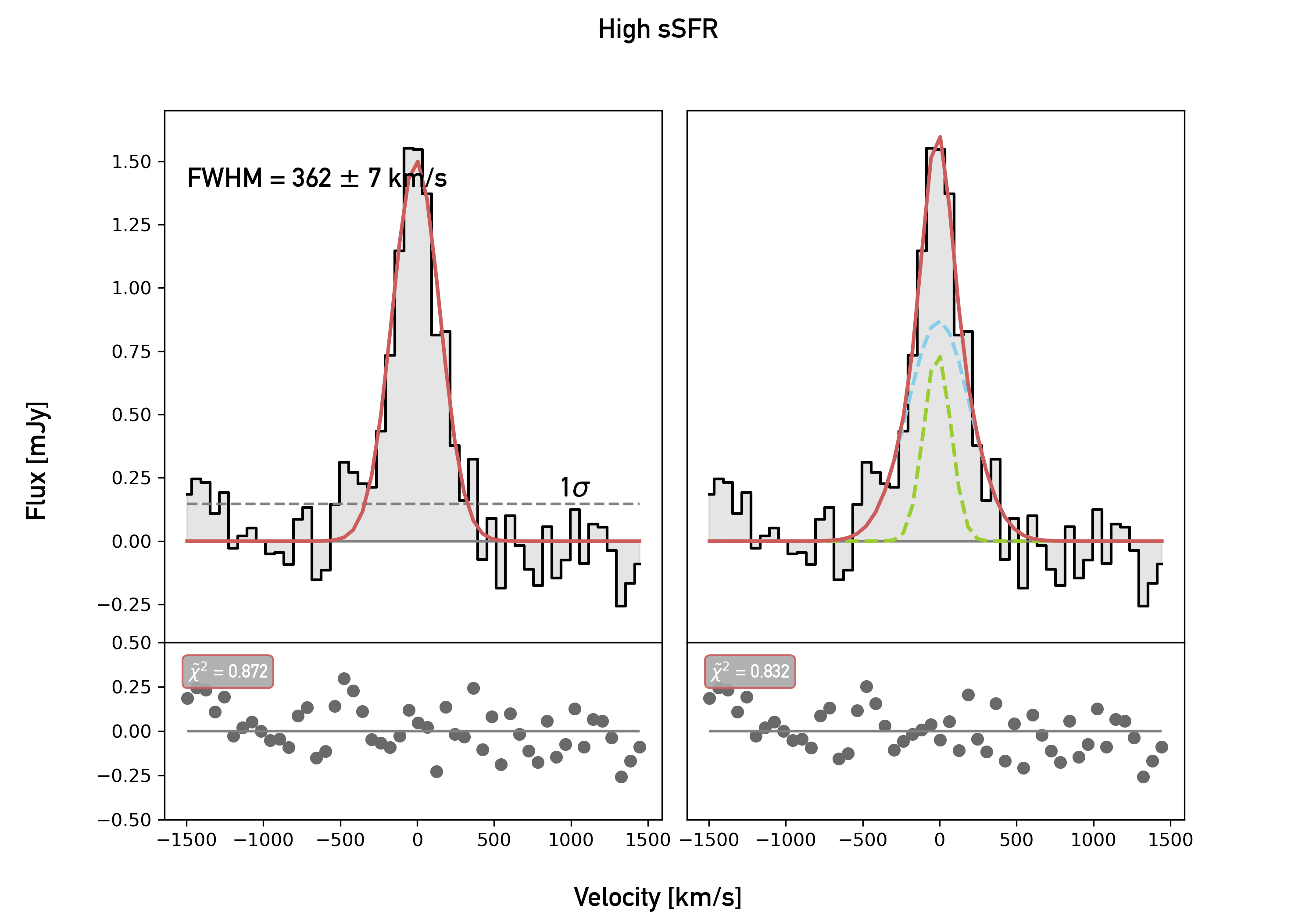}
    \end{subfigure}
    \begin{subfigure}{\columnwidth}
        \includegraphics[width=\linewidth]{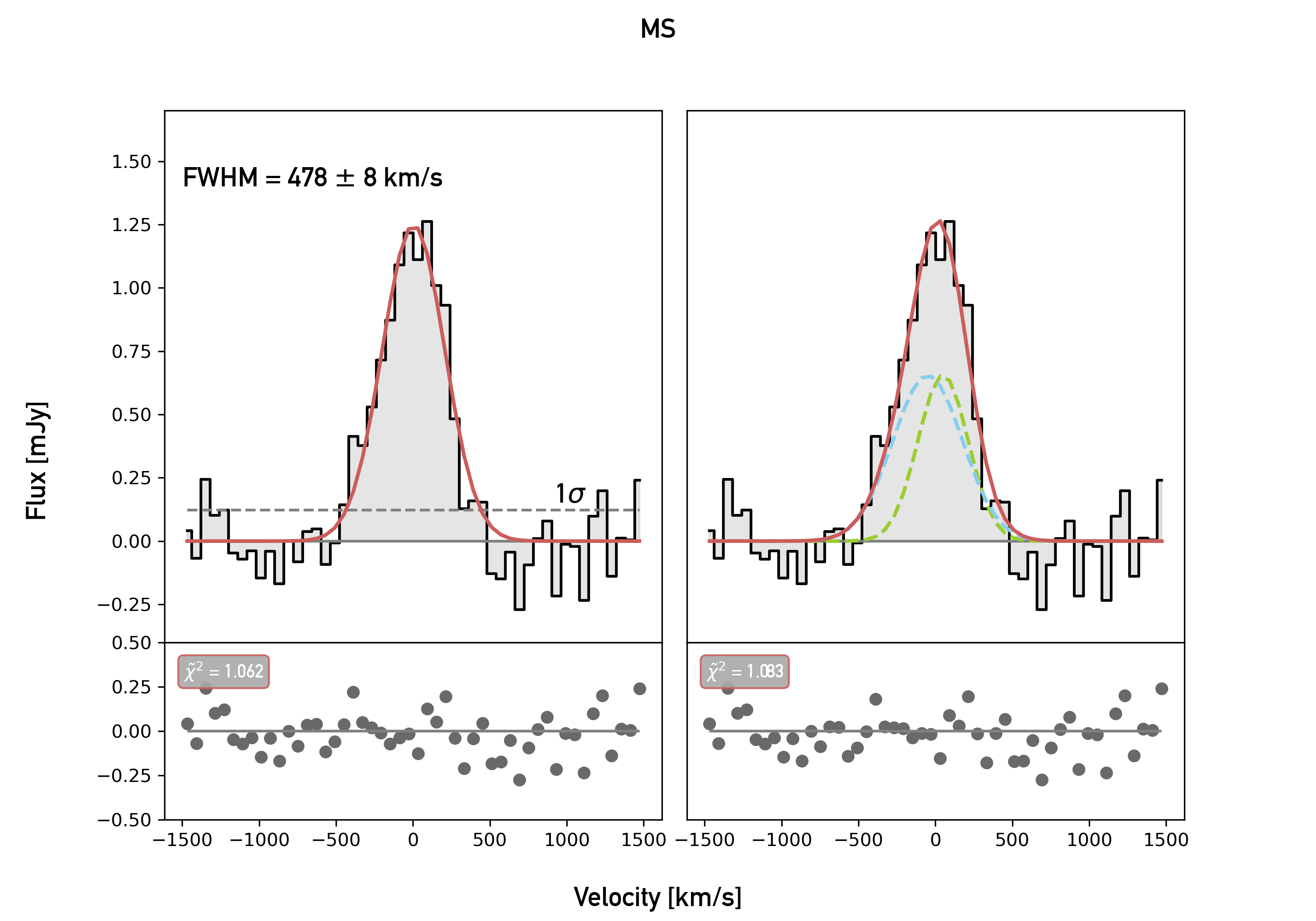}
    \end{subfigure}
    \begin{subfigure}{\columnwidth}
        \includegraphics[width=\linewidth]{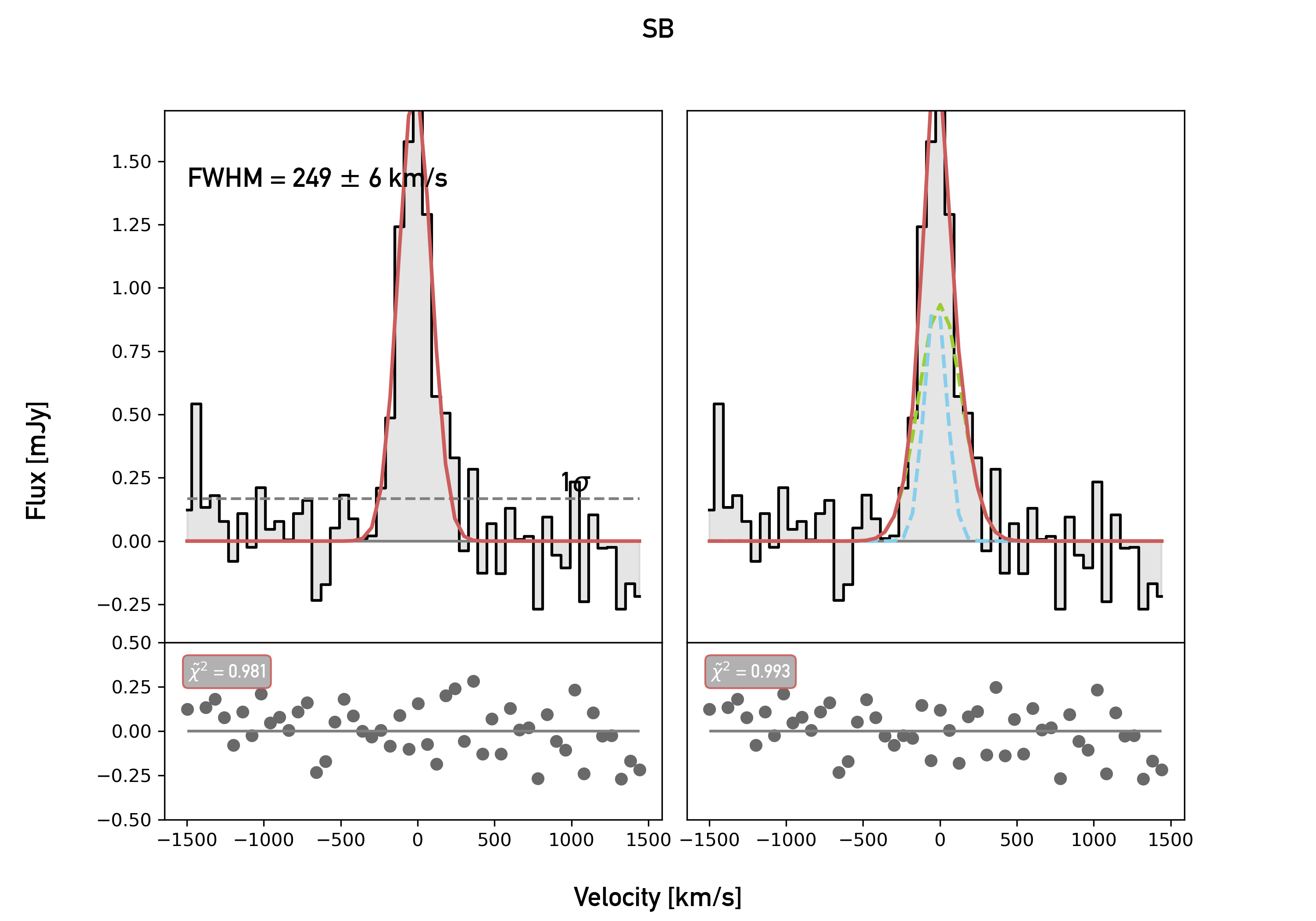}
    \end{subfigure}
    \caption{Stacks according to the properties of the galaxies. From top to bottom, we show the stellar mass bin, the infrared luminosity bin, the specific star formation rate bin, and the bin according to the position with regard to the MS. The first panel of the left and right columns shows single-Gaussian fits, while the second panel shows two-Gaussian fits.}
    \label{fig:binnedspectra}
\end{figure*}

We explore potential correlations between the properties of the galaxies in our sample and their stacked CO(2--1) emission line. We bin our sample in bins of $M_\star$, $L_{IR}$, sSFR and position with respect to the MS prior to stacking the spectra, as explained in section \ref{subsec:velocity}. We show the results in Figure \ref{fig:binnedspectra}. In all stacks, the single-Gaussian and two-Gaussian fits show similar residuals in their $\tilde{\chi}^2$, i.e., no broad component is required to improve the fit. However, the width of the CO(2--1) emission line varies across the different bins. The most striking case is in the MS and SB bins, where FWHM $= 478 \pm 27 \text{km\,s}^{-1}$ and FWHM$= 249 \pm 18 \text{km\,s}^{-1}$, respectively. The FWHM varies by almost a factor of 2. Similar to the full sample stack (Figure \ref{fig:SpectrumStack}), we note a weak asymmetry of the emission line with a marginal flux excess at negative velocities. This asymmetry is especially noticeable in the high $L_{IR}$ and high sSFR bins, where we find 3 consecutive channels (180kms$^{-1}$) with a a residual larger than the noise of the corresponding data, corresponding to a flux excess with a SNR of 3. This marginal flux excess might be a hint of outflows.

\section{Discussion}
\label{section:discussion}
\subsection{The spatial extent of the molecular gas of MS SFGs}
One piece of evidence for the capacity of outflows to entrain molecular gas to CGM scales is the presence of cold molecular gas on such scales (typically of the order of $\gtrsim 10$ kpc, well beyond the stellar emission). Through stacking 26 ALMA CO(2--1) intensity maps, corresponding to 26 MS SFGs at cosmic noon ($z_{med}=1.3$), we detect marginally resolved CO emission, when comparing the surface brightness radial profile of the stacked observations to the one of the stacked beam. This observed resolved CO emission extends to scales of $\sim 15$ kpc ($\sim1.9\arcsec$, at $z_{med}=1.3$) and the beam-corrected CO emission extends out to scales of $\sim12$ kpc ($1.3\arcsec$, see Section \ref{subsec:intensity}). A caveat to our analysis to note is that we did not perform PA alignment. This will be addressed in future work, however limited spatial resolution and dust obscuration could prevent accurate alignment and thus need to be addressed carefully. When repeating the stacking in bins of different galaxy properties (i.e., $M_\star$, $L_{IR}$, sSFR and $\Delta MS$), we find no differences in the extent of the CO emission within the different bins. However, we find marginally resolved CO emission in all bins. Our resampling test confirms that this extended CO detection is robust and representative of our sample of MS SFGs at cosmic noon.


\cite{Jones2023} recently found molecular gas, traced by a stack of ALMA CO(3--2) observations of AGN-host galaxies at cosmic noon, on scales of $\sim13$ kpc (observed scales). They also examined the extent of the stellar component, traced by stacking HST/i band observations of their sample, and found the stellar component (convolved with the synthesised beam) to not extend further than 10 kpc. Therefore, they concluded that the observed faint CO emission is distinct from the stellar component, tracing molecular gas around their galaxies, i.e., on CGM scales. The difference in the results found in \cite{Jones2023} and in our analysis on our sample of SFGs at cosmic noon could be interpreted as AGN-driven outflows being more effective at expelling molecular gas to large radii than star formation-driven outflows.


\cite{Puglisi2019} show a comparison between sizes traced by Ks band observations, i.e., stellar sizes, and sizes traced by multiple CO transition and/or dust-continuum observations of MS SFGs at the same redshift of our work, including our sample. They find that the "ALMA size" (corresponding to the effective radius of multiple CO, [CI] and dust-continuum observations) does not exceed $r_e \sim7$ kpc (mostly upper limits) at all stellar masses, and is more compact than the sizes traced by the Ks band observations (convolved with the synthesised beam of their observations). Our results on stacked CO(2--1) observations show that the half-light radius\footnote{for a Gaussian profile, the Sersic half-light radius and effective radius are equivalent} of the cold molecular gas emission is $r_e \sim6$ kpc, consistent with the results of \cite{Puglisi2019}. Indeed, our Ks band stack to ALMA CO stack comparison (Figure \ref{fig:ALMAKs}) shows that the two corresponding profiles follow each other, indicating no CO emission beyond the stellar emission. Simulations (e.g., \citealt{Bahe2016}, \citealt{Pillepich2019}) show stellar and ionised gas components of MS galaxies at cosmic noon extending to $\sim10$ kpc, with sizes varying with stellar mass. However, the spatial extent of CO emission in simulations remains mostly unexplored, leaving a gap in comparison with observations. In conclusion, our comparison of ALMA CO emission and UltraVISTA Ks band data reveals molecular gas distributed on large physical scales, with no clear evidence of these scales extending beyond the stellar discs of our sample of cosmic noon MS SFGs, in agreement with recent findings in \citet{Howatson2025}.

While this conclusion stands, individual visual inspection of the Ks band images sometimes showed disturbed morphologies, that could hint towards satellite galaxies or recent mergers contributing to stellar emission on large scales. High-resolution integral field spectroscopy observations are needed to test whether the large-scale stellar emission is dynamically associated with the central galaxies or driven by faint satellites and mergers.

\subsection{The molecular gas kinematics of MS SFGs}
Another piece of evidence for outflows entraining cold molecular gas to the CGM would be observations of low-level high velocity (broad wings) CO emission. We note that finding broad wings in the CO kinematics does not necessarily imply molecular gas surrounding galaxies, hence our investigation for spatially extended CO emission discussed in the previous paragraph. In Section \ref{subsec:velocity}, we show no clear detection of broad wings in the stack spectrum (Figure \ref{fig:SpectrumStack}). \citet{Barfety2025} stack 154 typical massive SFGs at $z=0.5-2.6$ with different CO observation (CO(3--2), CO(2--1) and CO(6--5)), and found similar results, i.e., no clear outflow signatures are found when stacking the spectra of their full sample. Nevertheless, the lack of detected broad wings does not rule out the presence of outflows all together. It rules out the presence of prominent fast ongoing outflows. Indeed, several studies have shown that molecular outflows in extreme objects such as ultra luminous infrared galaxies (ULIRGs) or quasars can be several tens of times fainter than the bulk of the emission from the central source (e.g., \citealt{Feruglio2010}, \citealt{Cicone2012}, \citealt{Feruglio2015}, \citealt{PereiraSantaella2016}, \citealt{Veilleux2017}, \citealt{PereiraSantaella2018}, \citealt{Saito2018}, \citealt{Brusa2018}, \citealt{Bischetti2019a}, \citealt{Cicone2020}). While our sample of galaxies represent a different population of galaxies, the lack of detected broad wings might reflect unsuficient observational depth.

In nearly all the stacked spectra (i.e., stacking the entire sample or bins), we note the presence of a tentative excess of CO(2--1) emission at negative velocities. This asymmetry could suggest the presence of outflows, with the receding part not visible due to optical depth effect caused by dense CO gas in galaxies. This interpretation is further supported by the stronger separation observed between the emission at the galaxy systemic velocity and the emission at negative velocities observed in the stacked high-infrared luminosity and high sSFR bins (suggestive of high dust and gas content), corresponding to more recent star-formation that could produce stellar-driven outflows. These results and interpretation corroborate with the ones presented in \citet{Bischetti2019b}, where an excess of [CII] emission at negative velocities is found for a stack of 48 quasars $z\sim5.5$.

Outflows are commonly detected in their ionised phase, typically in the form of broad wings present in H$\alpha$ emission out to $z\sim2.5$ in SFGs (e.g., \citealt{ForsterSchreiber2019}, \citealt{RodriguezdelPino2019}). Multi-phase, including the cold phase, outflows were also detected in typical SFGs in the literature using [CII] emission (e.g., \citealt{Ginolfi2020}). Therefore, we might have expected to detect them in our sample as well. If not due to the lack of observational depth, the lack of observed prominent broad wings in our stack of CO emission of MS SFGs suggest the different phases of outflows undergo different physical mechanisms (e.g., \citealt{Toba2017}).

\subsection{A brief contrast with \textit{uv}-plane analysis}
We have conducted our analysis in the image-plane. Loss of information can occur or nonphysical artefact can be induced when imaging interferometric data. A \textit{uv}-plane analysis (e.g., \citealt{Vio2017}, \citealt{Vio2019}, \citealt{Tsukui2022}, \citealt{Tsukui2023}) would provide a complementary approach to test the robustness of the results presented here. For instance, \cite{Rybak2024} very recently conducted a uv-plane stacking analysis of Early Karl. G. Jansky Very Large Array (JVLA) CO(1--0) observations of a sample of 19 dusty SFGs at $z=2.0-4.5$. Their stacked CO(1--0) spectrum is best fitted by a single Gaussian profile, revealing the absence of broad wings, albeit a positive excess of flux at negative velocities. While the results of our analysis on the CO(2--1) stacked spectrum of our sample are consistent with the results presented in \cite{Rybak2024}, our results regarding the spatial extent of the CO emissions differ. The uv-plane stack in \cite{Rybak2024} reveals CO(1--0) emission out to $\sim4$ kpc (half-light radius), beyond the stellar emission as traced by \textit{James Webb Space Telescope} NIRCam $4.4 \mu$m imaging. Compared to our sample, the sample in \cite{Rybak2024} consists of higher-redshifts, more star-forming and dusty galaxies. These properties can have an impact on the capacity of outflows in driving molecular gas outside galaxies.

\section{Summary and outlook}
\label{subsec:summaryandoutlook_stacking}
In this work we have presented a stacking analysis of the CO(2--1) emission of 26 galaxies on and above the MS at cosmic noon, first presented in \cite{Valentino2020}. Our goal in this work was to understand if outflows in typical cosmic noon SFGs are capable of ejecting molecular gas into the CGM. To achieve this goal, our means were two-fold. 1) We aimed at quantifying the spatial extent of the molecular gas. For this purpose we stacked the ALMA archival CO(2--1) intensity maps of our sample. 2) We searched for the presence of ongoing outflows in our sample, in the form of low-brightness high velocity molecular gas. To this end, we stacked the CO(2--1) emission spectra and performed two-Gaussian fits to reveal the presence (or absence) of broad wings.

\begin{itemize}
    \item Stacking 26 individual CO(2--1) intensity maps and computing the corresponding surface brightness radial profiles (i.e., of the stack CO(2--1) emission and synthesised beam), we find marginally resolved faint CO(2--1) emission. The beam-resolved faint emission extends out to $1.9\arcsec$ ($\sim16$ kpc), with a median beam of $1.5\arcsec$ for our observations. Our Gaussian beam-correction shows intrinsic CO(2--1) emission present out to $1.4\arcsec$ ($\sim12$ kpc).
    \item Resampling the intensity map stack and corresponding radial profiles shows that the marginally resolved faint emission is a ubiquitous feature of our sample of MS SFGs.
    \item No significant difference in the radial profile is found when separating the stacks in bins of low and high $M_\star$, $L_{IR}$, and sSFR, respectively and bins of MS versus starburst galaxies.
    \item Comparing the intrinsic CO(2--1) stacked radial profile to the one of the deconvolved Ks band stack reveals coincidental emissions, with both emissions present out to $\sim12-13$ kpc. Therefore, there is no evidence for CO emission significantly extending beyond the stellar emission.
    \item Stacking 26 individual CO(2--1) emission line spectra, we do not observe the presence of prominent molecular outflows traced by broad wings.
    \item When performing binned stacks of the CO(2--1) emission line spectra according to different galaxy properties, we do not observe the presence of prominent molecular outflows traced by broad wings.
    \item We note the presence of a tentative "blue bump', i.e., an excess of flux at negative velocities, in the stacked spectrum of all 26 individual spectra and nearly all binned spectra. This "blue bump" could be indicative of outflows, with the redshifted part being optically thick.
\end{itemize}

In this study, although we found no evidence for molecular outflows, our stacking analysis has revealed compelling evidence of CO molecular emission extending over large spatial scales, exceeding 10 kpc. To accurately quantify the full extent of CO emission and explore its correlation with the stellar and star-formation properties of SFGs, deep ALMA observations of individual sources will be essential. These observations will yield more precise measurements, which, when combined with state-of-the-art JWST observations for instance, will enable a comprehensive understanding of the relationship between the molecular gas and the star formation content of SFGs, ultimately shedding light on broader galaxy evolution processes. At the same time, mapping the diffuse, low–surface-brightness CGM on such scales, where molecular outflows might have entrained gas, will require facilities beyond ALMA’s interferometric capabilities. In this context, AtLAST, with its large collecting area and wide field of view, will be uniquely suited to capturing the extended emission missed by interferometers, providing a vital complement to ALMA and JWST in connecting galaxies with their surrounding gaseous haloes \citet{Schimek2024}.


\begin{acknowledgements}
The work presented in this paper makes use of ALMA data ADS/JAO.ALMA\#2016.1.00171.S. ALMA is a partnership of ESO (representing its member states), NSF (USA) and NINS (japan), together with NRC (Canada), MOST and ASIAA (Taiwan), and KASI (Republic of Korea), in cooperation with the Republic of Chile. We thank the anonymous referee for their feedback that allowed us to improve the clarity of the manuscript. We also thank Melanie Kaasinen, Joshiwa van Marrewijk, Santiago Arribas, Bruno Rodríguez del Pino and Román Fernández Aranda for helpful and insightful discussions. GCJ acknowledges support by the Science and Technology Facilities Council (STFC), by the ERC through Advanced Grant 695671 ``QUENCH'', and by the UKRI Frontier Research grant ``RISEandFALL.'' Based on data products from observations made with ESO Telescopes at the La Silla Paranal Observatory under ESO programme ID 179.A-2005 and on data products produced by TERAPIX and the Cambridge Astronomy Survey Unit on behalf of the UltraVISTA consortium.
      
\end{acknowledgements}

%
%
\bibliographystyle{aa} 
\bibliography{biblio.bib} 






   
  




\begin{appendix} 
\onecolumn
\section{Additional material}
\label{section:apd}

\begin{table*}[ht!]
\centering
\begin{tabular}{|c|c|c|c|c|c|c|c|}
\hline
\textbf{ID} & \textbf{RA} & \textbf{DEC} & \textbf{Line sens. 10km$/$s} & \textbf{Restoring beam} & \textbf{Observing time}& \textbf{Angular resolution} \\
 & & & [mJy$/$beam] & [arcsec] & [s] & [arcsec] \\
\hline
2299 & 10h02m15.36s & +01d54m5.33s &1.784 & 1.6$\times$1.17 & 725.760 & 1.054 \\

2993 & 10h00m2.07s & +02d01m32.7s &2.445 & 1.6$\times$1.4 & 272.160 & 1.066 \\

13205 & 10h02m17.96s & +01d56m52.3s &2.451 & 1.6$\times$1.4 & 272.160 & 1.114 \\

13854 & 10h02m32.91s & +02d00m27.61s &2.451 & 1.6$\times$1.4 & 272.160 & 1.114 \\

15069 & 09h58m47.07s & 02d07m21.43s &1.890 & 1.8$\times$1.4 & 483.840 & 1.138 \\

18911 & 10h00m11.55s & +02d28m3.77s &2.445 & 1.6$\times$1.4 & 272.160 & 1.070 \\

19021 & 10h00m14.11s & +02d28m38.6s &1.890 & 1.8$\times$1.4 & 483.840 & 1.141 \\

21060 & 10h00m29.94s & +02d40m7.03s &2.451 & 1.6$\times$1.4 & 272.160 & 1.116 \\

21820 & 10h02m32.69s & +02d44m39.72s &1.784 & 1.6$\times$1.2 & 725.750 & 1.058 \\

26925 & 10h01m51.24s & +01d52m59.92s &2.445 & 1.6$\times$1.4 & 272.160 & 1.067 \\

27172 & 10h01m40.24s & +01d55m49.66s &1.890 & 1.8$\times$1.4 & 483.840 & 1.136 \\

30122 & 10h00m8.76s & +02d19m1.96s &2.445 & 1.6$\times$1.4 & 272.160 & 1.065 \\

30694 & 09h58m38.46s & +02d24m39.56s &2.445 & 1.6$\times$1.4 & 272.160 & 1.075 \\

31880 & 10h00m50.46s & +02d33m55.51s &1.784 & 1.6$\times$1.2 & 725.760 & 1.058 \\

34023 & 09h59m51.31s & +02d52m48.86s &1.784 & 1.6$\times$1.2 & 725.760 & 1.057 \\

35349 & 10h01m59.76s & +01d43m28.06s &1.890 & 1.7$\times$1.3 & 483.840 & 1.132 \\

37250 & 09h58m28.35 & +02d11m36.46s &2.445 & 1.6$\times$1.4 & 272.160 & 1.081 \\

38053 & 10h02m13.64s & +02d23m29.04s &2.445 & 1.6$\times$1.4 & 272.160 & 1.075 \\

41210 & 09h59m31.48s & +02d27m2.52s &2.451 & 1.6$\times$1.4 & 272.160 & 1.117 \\

41458 & 10h00m53.56s & +02d41m25.37s &2.451 & 1.6$\times$1.4 & 272.160 & 1.116 \\

42925 & 10h01m53.27s & +01d50m37.97s &2.451 & 1.6$\times$1.4 & 272.160 & 1.114 \\

44641 & 10h02m38.05s & +02d19m4.29s &2.445 & 1.6$\times$1.4 & 272.160 & 1.079 \\

44694 & 10h01m36.15s & +02d20m3.98s &1.890 & 1.8$\times$1.4 & 483.840 & 1.138 \\

48881 & 09h59m8.22s & +01d54m45.79s &2.445 & 1.6$\times$1.4 & 272.160 & 1.077 \\
 
51330 & 09h59m57.97s & +01d43m27.26s &2.451 & 1.7$\times$1.4 & 272.160 & 1.113 \\

51936 & 09h59m24.46s & +02d43m6.92s &1.784 & 1.6$\times$1.2 & 725.760 & 1.058 \\
\hline
\end{tabular}
\caption{Table reporting the technical information about the interferometric data used this paper, as reported in \citealt{Valentino2020}. For all data, the observations were taken in array configuration C40-4.}
\label{table:technical}
\end{table*}

\begin{table*}[ht!]
\centering
\begin{tabular}{|c|c|c|c|c|c|}
\hline
\textbf{ID} & \textbf{Redshift} & \textbf{Log Stellar Mass [M$_\odot$]} & \textbf{Infrared Luminosity [$10^{12}$ L$_\odot$]} & \textbf{Log SFR [M$_\odot$ yr$^{-1}$]} & \textbf{SNR CO(2--1)} \\
\hline
2299$^\dag$ & $1.3948\pm0.00025$ & 10.8221 & $4.911\pm 0.06045$ & 2.864 & 15 \\

2993 & $1.1917\pm0.00031$ & 10.9888 & $1.997\pm0.1152$ & 2.473 & 7 \\

13205 & $1.2659\pm0.00038$ & 11.1442 & $1.619\pm0.1395$ & 2.382 & 5 \\

13854 & $1.2695\pm0.00024$ & 11.0327 & $2.260\pm0.1609$ & 2.527 & 7 \\

15069 & $1.2119\pm0.00025$ & 10.61 & $1.213\pm 0.09902$ & 2.257 & 4 \\

18911 & $1.1710\pm0.00023$ & 10.6379 & $0.801\pm 0.06477$ & 2.077 & 3 \\

19021 & $1.2581\pm0.00024$ & 10.2025 & $3.276\pm 0.09806$ & 2.688 & 8 \\

21060 & $1.2818\pm0.00024$ & 9.8400 & $1.834\pm0.1005$ & 2.436 & 4 \\

21820$^\dag$ & $1.3790\pm0.00032$ & 10.9209 & $1.863\pm0.06887$ & 2.443 & 6 \\

26925 & $1.1671\pm0.00030$ & 11.2428 & $0.842\pm0.07139$ & 2.098 & 7 \\

27172 & $1.2348\pm0.00032$ & 10.757 & $1.002\pm0.09275$ & 2.174 & 4 \\

30122 & $1.4626\pm0.00031$ & 11.08 & $2.086\pm0.09912$ & 2.492 & 9 \\

30694 & $1.1611\pm0.00021$ & 10.7949 & $0.877\pm0.09839$ & 2.116 & 8 \\

31880$^\dag$ & $1.3999\pm0.00034$ & 11.0176 & $1.888\pm0.06537$ & 2.449 & 7 \\

34023$^{\dag,\wedge}$ & $1.3461\pm0.00032$ & 10.97 & $1.694\pm0.09633$ & 2.402 & 3 \\

35349 & $1.2553\pm0.00045$ & 11.1876 & $1.221\pm0.03919$ & 2.259 & 8 \\

37250 & $1.1525\pm0.00034$ & 10.95 & $1.831\pm0.08732$ & 2.436 & 14 \\

38053 & $1.1550\pm0.00036$ & 10.7301 & $1.074\pm0.1345$ & 2.204 & 6 \\

41210 & $1.3145\pm0.00021$ & 10.5562 & $2.580\pm0.1015$ & 2.585 & 12 \\

41458 & $1.2949\pm0.00022$ & 10.8348 & $3.083\pm0.1544$ & 2.662 & 9 \\

42925 & $1.6028\pm0.00054$ & 9.4012 & $4.900\pm0.2856$ & 2.863 & 6 \\

44641 & $1.1509\pm0.00024$ & 11.55 & $1.054\pm0.04951$ & 2.196 & 5 \\

44694 & $1.2056\pm0.00036$ & 10.7568 & $0.776\pm0.1006$ & 2.063 & 6 \\

48881 & $1.1569\pm0.00022$ & 10.86 & $1.641\pm0.1714$ & 2.388 & 5 \\
 
51330 & $1.6317\pm0.00041$ & 10.2205 & $4.675\pm0.4038$ & 2.843 & 6 \\

51936$^\dag$ & $1.4030\pm0.00029$ & 10.243 & $2.034\pm0.1167$ & 2.481 & 9 \\
\hline
\end{tabular}
\caption{Table summarizing key properties of the sample used in this paper, as provided in \cite{Valentino2020}. $^\dag$ indicates the five galaxies where only one measurement set was kept to preserve beam consistency througout the sample. $^{\wedge}$Ultravista Ks band data is missing for this galaxy.}
\label{table:sample}
\end{table*}

\begin{figure*}[h]
    \centering
    \includegraphics[width=0.8\textwidth]{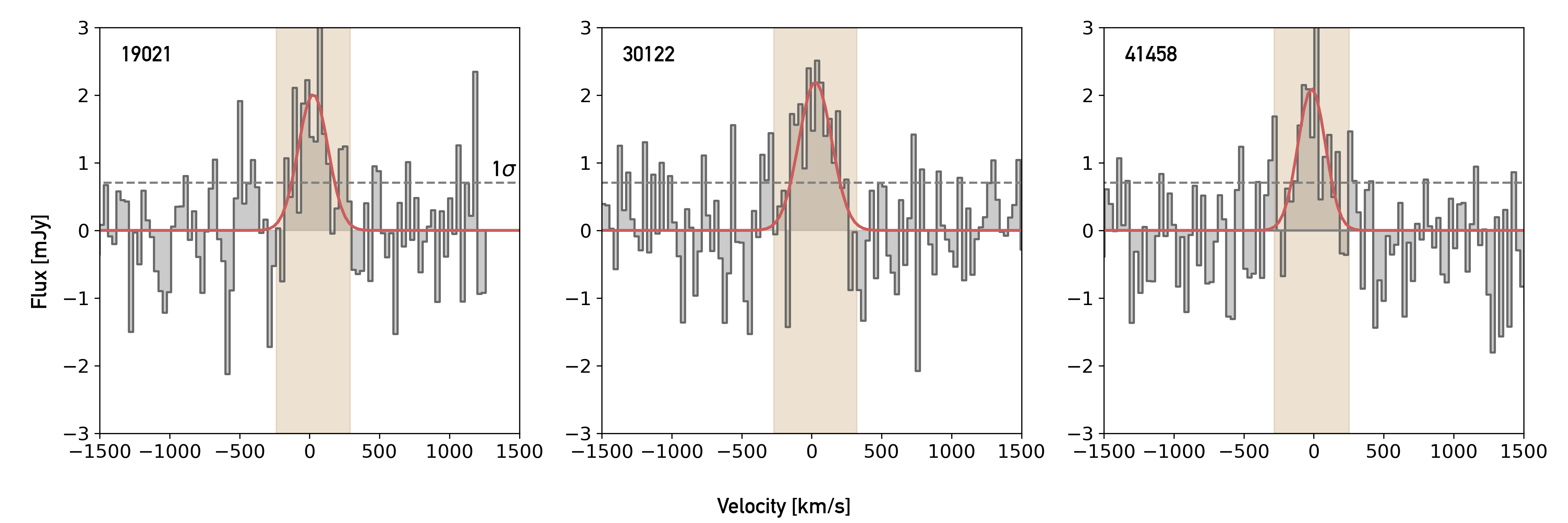}
    \caption{Examples of spectra (grey) and the median Gaussian fit (red solid line) used to image the intensity maps in section \ref{subsec:intensity}. The channels used to create the intensity maps (i.e., FWHM $\pm \sigma_{\text{FWHM}}$ of the median Gaussian fit) are highlighted with the light-brown shaded area. The horizontal dashed line shows the 1$\sigma$ noise level of the spectrum.}
    \label{fig:examplefits}
\end{figure*}

\begin{figure*}[h]
    \centering
    \begin{subfigure}[b]{0.4\textwidth}
         \centering
         \includegraphics[width=\textwidth]{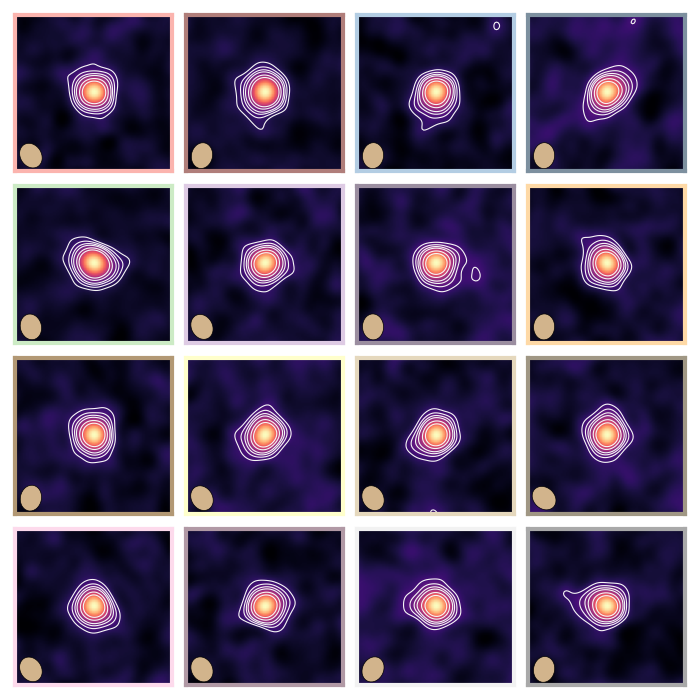}
         \label{fig:restackingmaps}
     \end{subfigure}
     \begin{subfigure}[b]{0.5\textwidth}
         \centering
         \includegraphics[width=\textwidth]{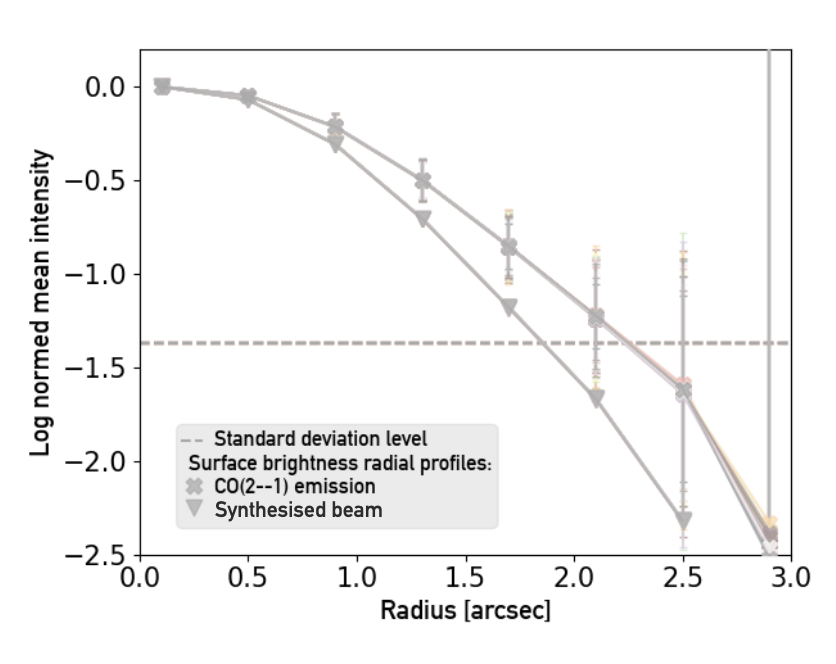}
         \vspace{8pt}
         \label{fig:restackingprofile}
     \end{subfigure}
\caption{Left: CO(2--1) intensity maps resulting from uniformly stacking the individual CO(2--1) intensity maps with random rotation. Contours at [3, 5, 7, 9, 13, 17]$\sigma$ are shown with white contours. The stacked beam is shown in the bottom left corner of each map. Each map is $5\times5$ arcsec. Right: Surface brightness radial profiles corresponding to each intensity map shown on the left, with matching their frame colors. Crosses represent the CO(2--1) radial profile and triangles represent the corresponding synthesised beam profile.}
\label{fig:restacking}
\end{figure*}

\begin{figure*}[h]
    \includegraphics[width=\textwidth]{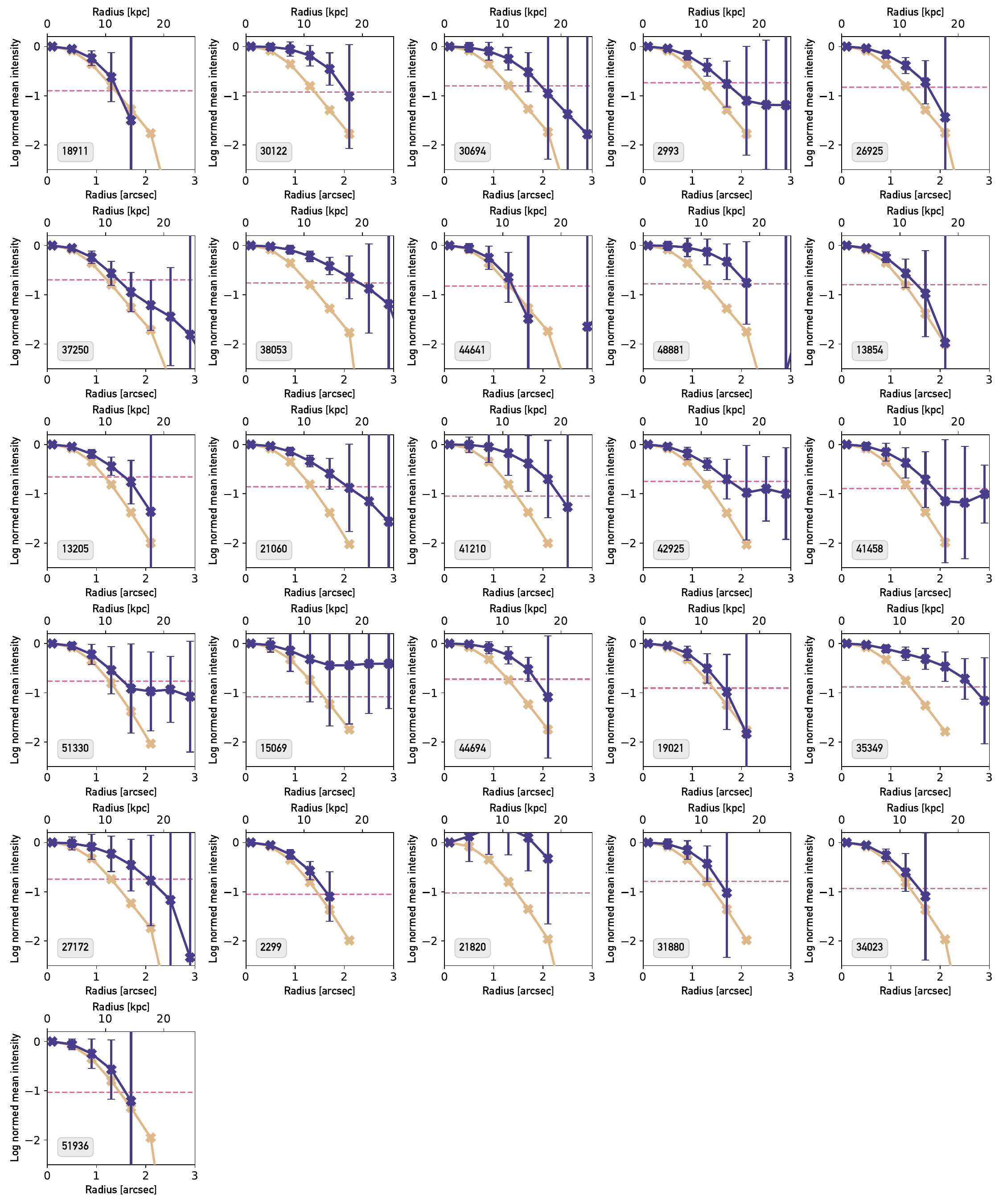}
        \caption{Stamps showing the surface brightness radial profile of each galaxy in our sample. The galaxy ID number is shown at the bottom left corner of each stamp. The colour-code is the same as in the main text, i.e. the horizontal pink dashed-line shows the $1 \sigma$ standard deviation level of the map, the beige solid line shows the radial profile of the synthesised beam and the purple solid line shows the radial profile of the CO(2--1) emission.}
        \label{stamps}
\end{figure*}
\end{appendix}

\begin{figure*}[h]
    \centering
    \begin{subfigure}{0.45\textwidth}
        \includegraphics[width=\linewidth]{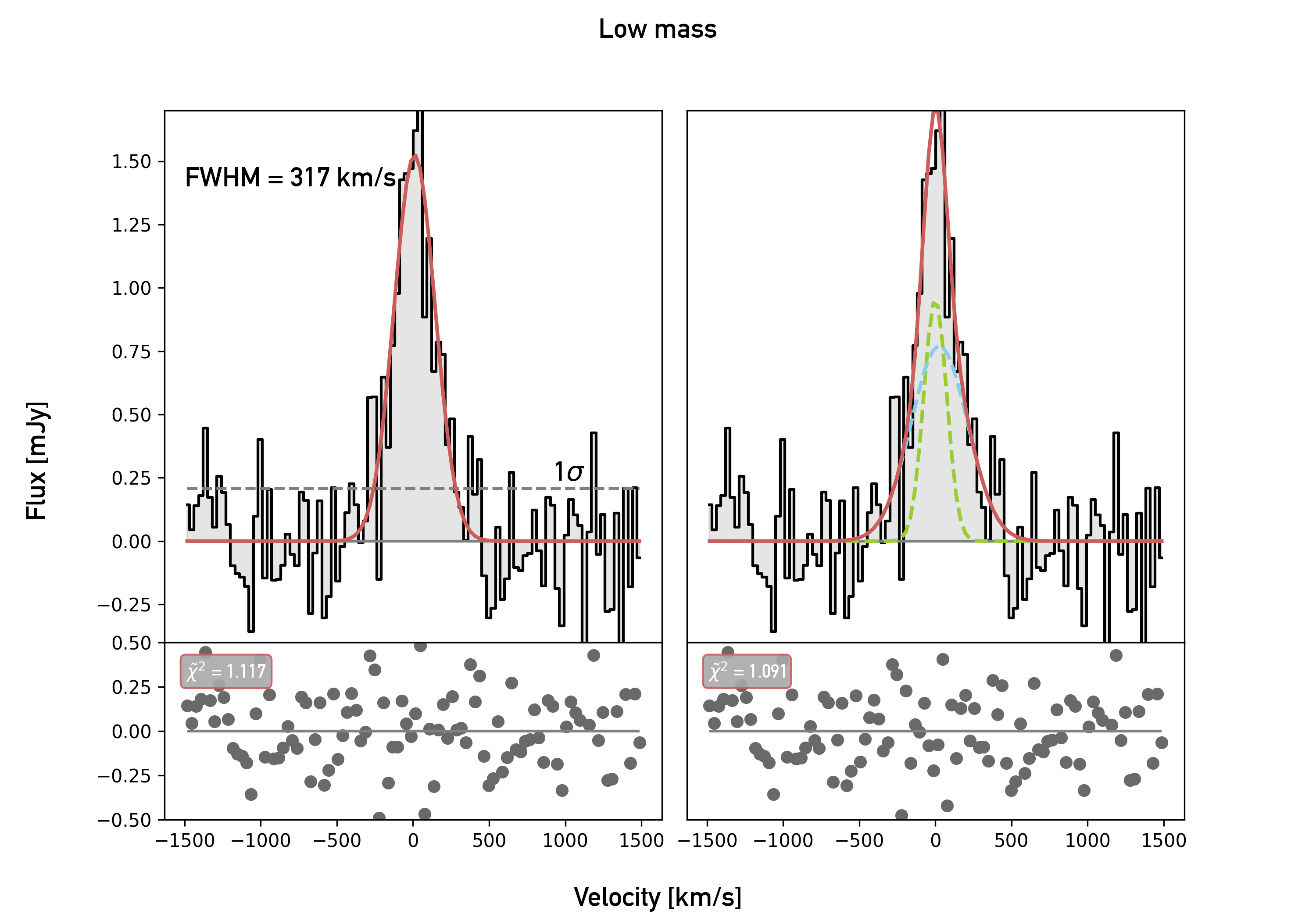}
        \label{fig:plot1}
    \end{subfigure}
    \begin{subfigure}{0.45\textwidth}
        \includegraphics[width=\linewidth]{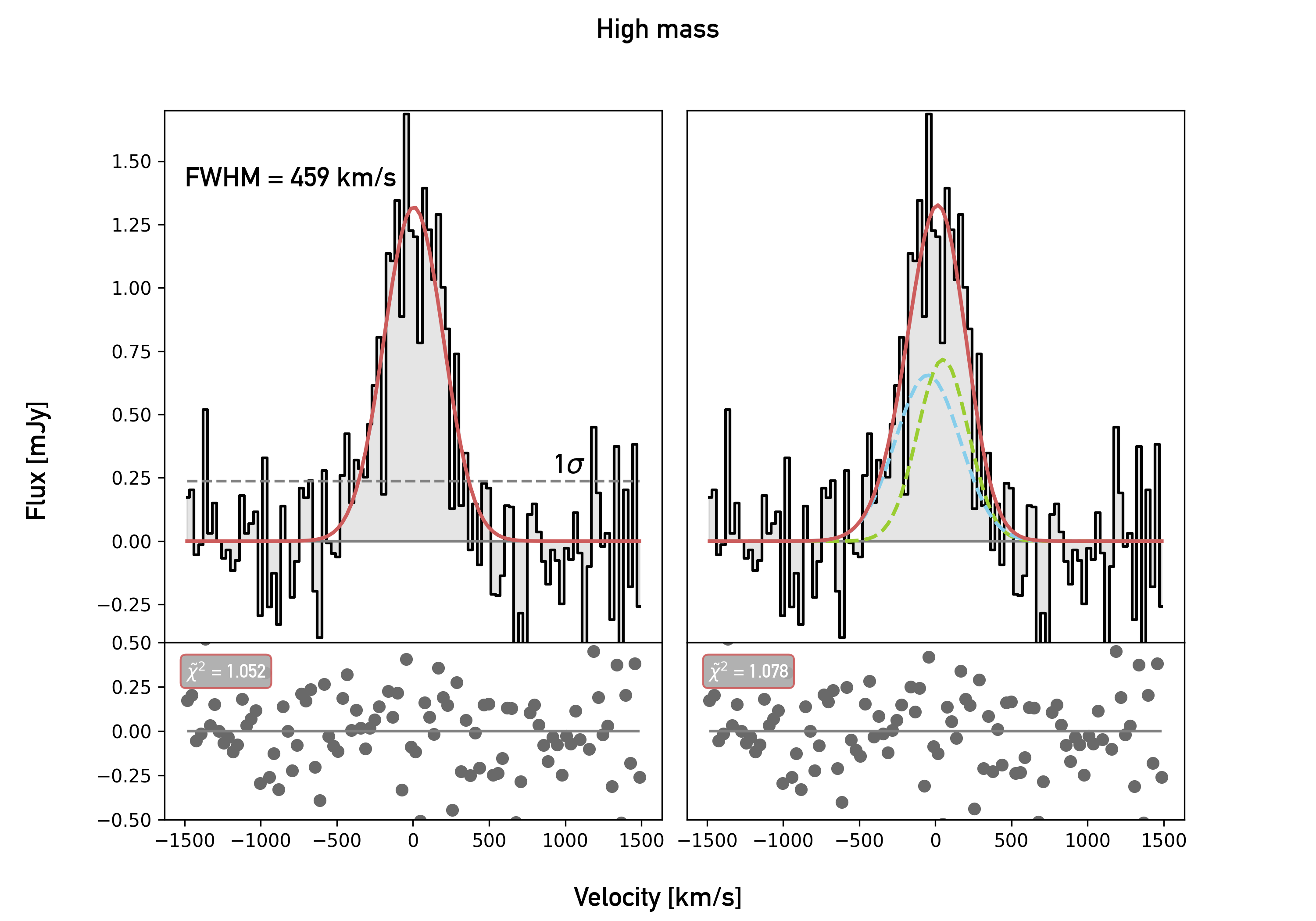}
        \label{fig:plot2}
    \end{subfigure}
    \begin{subfigure}{0.45\textwidth}
        \includegraphics[width=\linewidth]{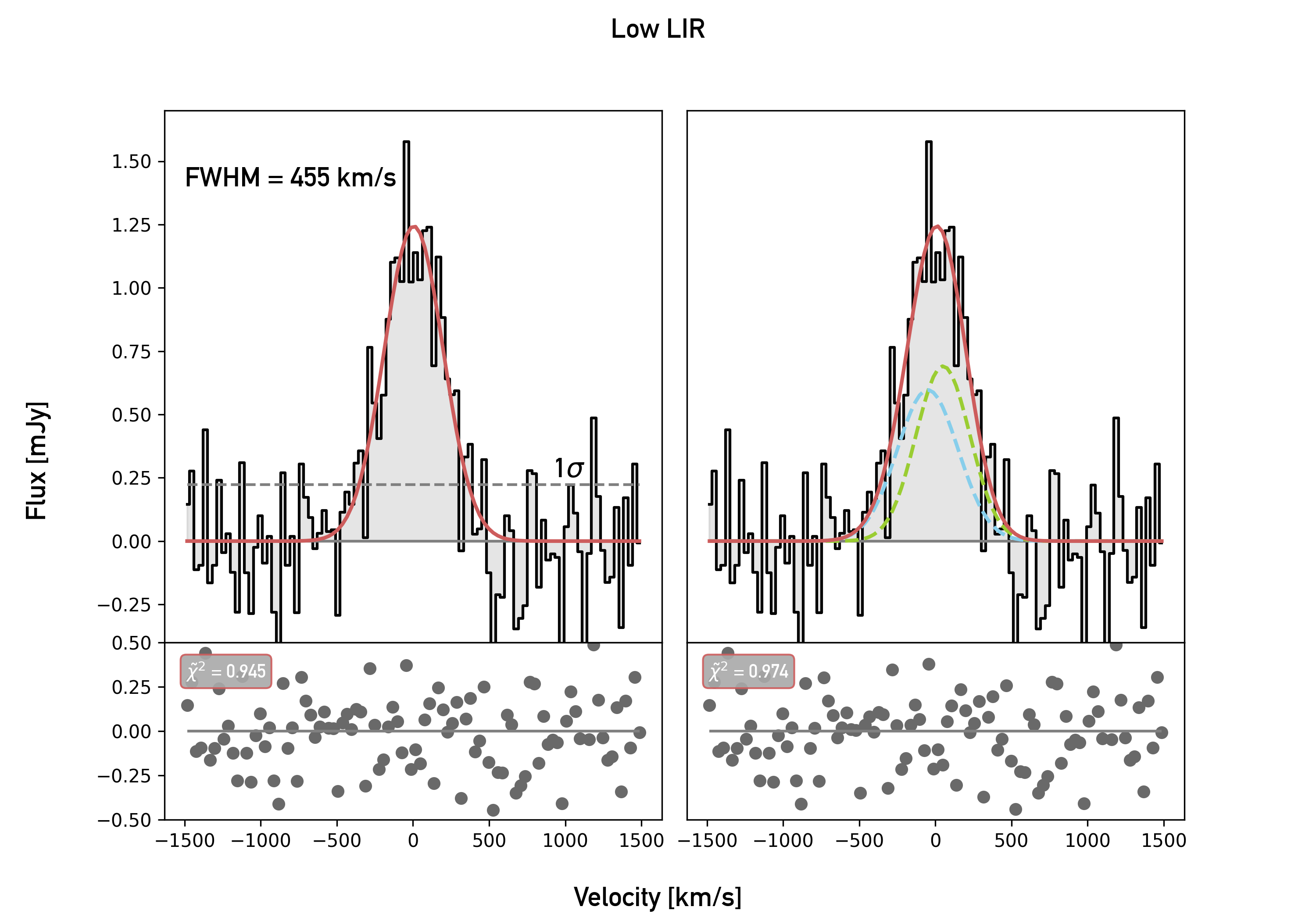}
        \label{fig:plot3}
    \end{subfigure}
    \begin{subfigure}{0.45\textwidth}
        \includegraphics[width=\linewidth]{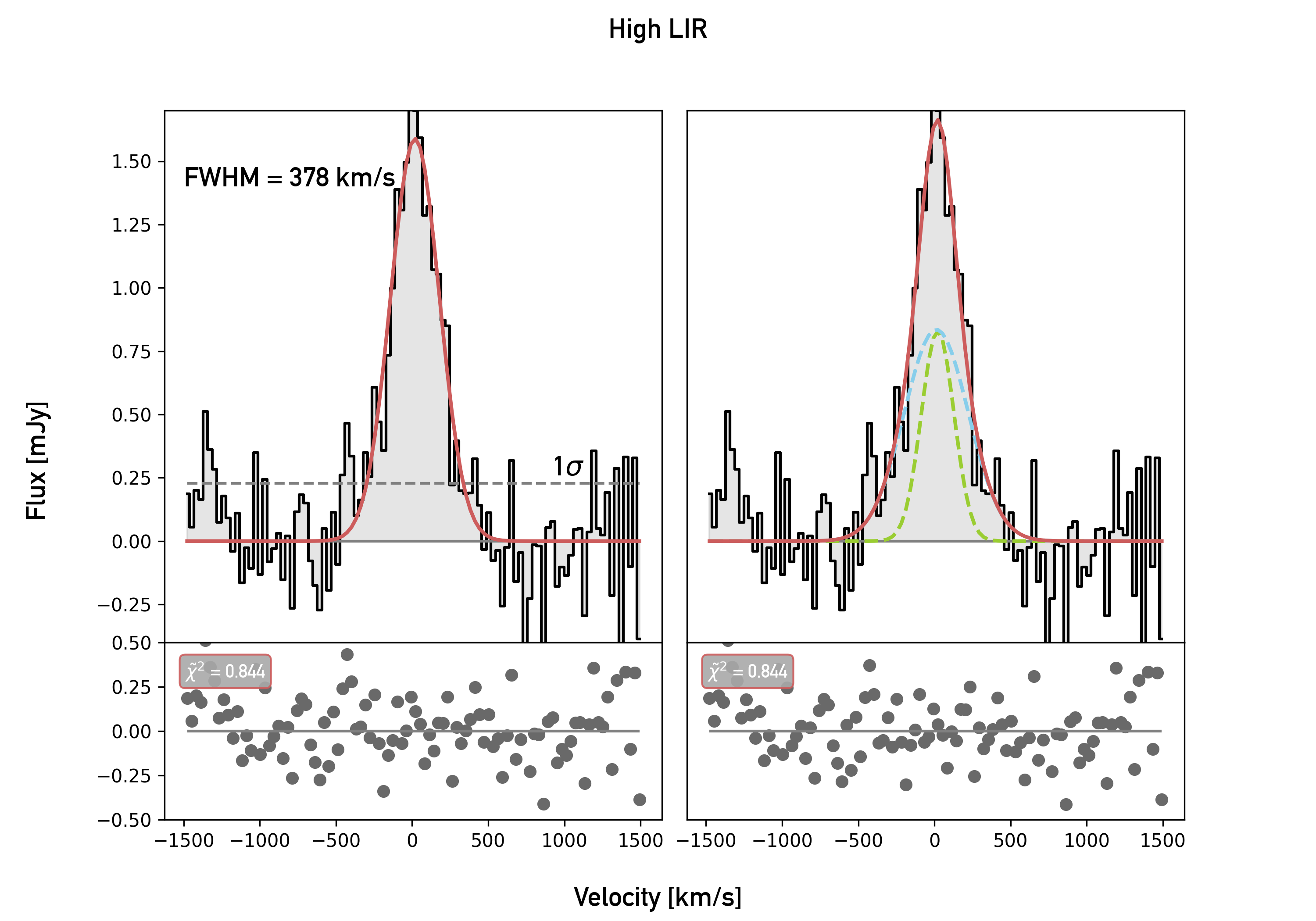}
        \label{fig:plot4}
    \end{subfigure}
    \begin{subfigure}{0.45\textwidth}
        \includegraphics[width=\linewidth]{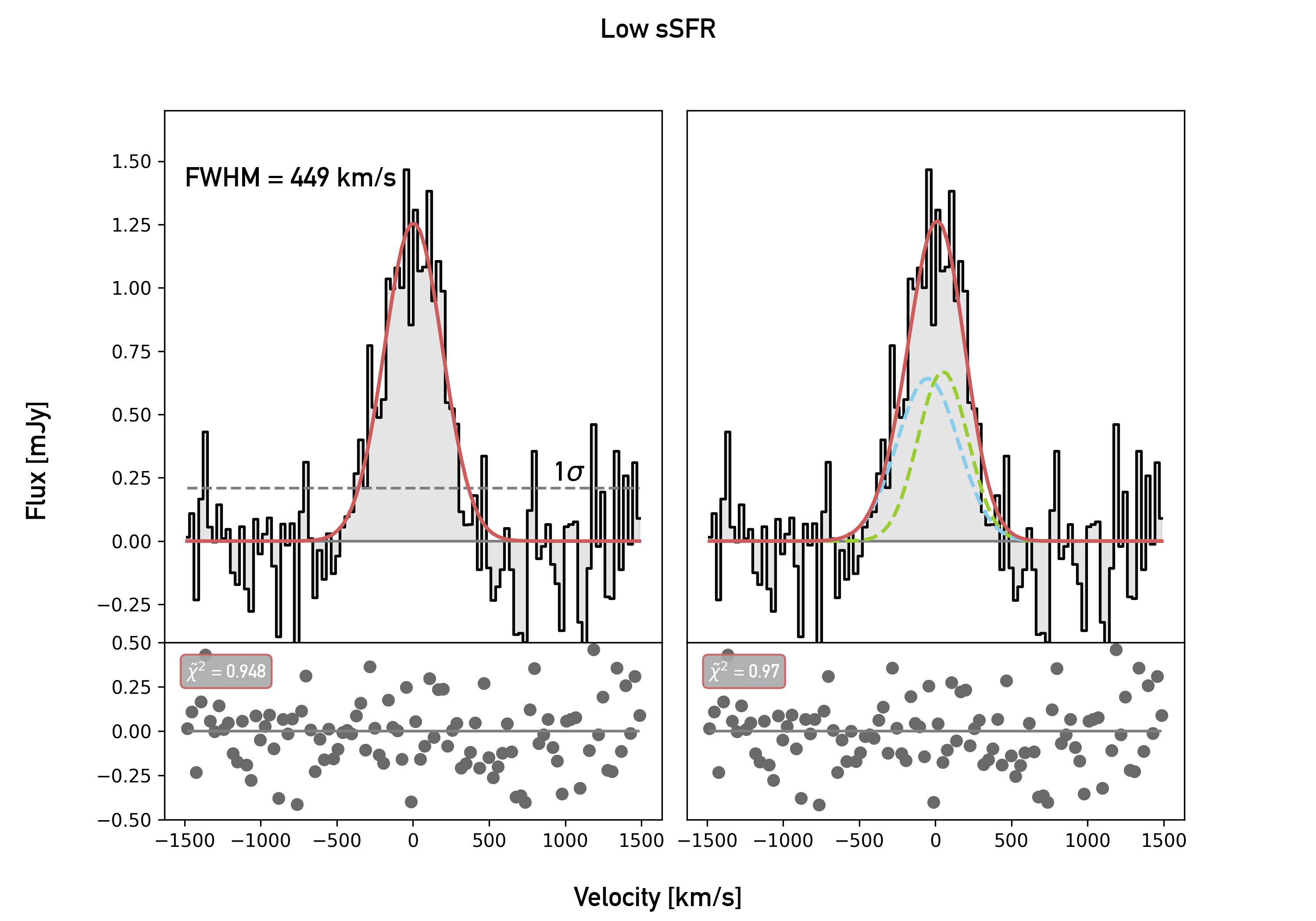}
        \label{fig:plot5}
    \end{subfigure}
    \begin{subfigure}{0.45\textwidth}
        \includegraphics[width=\linewidth]{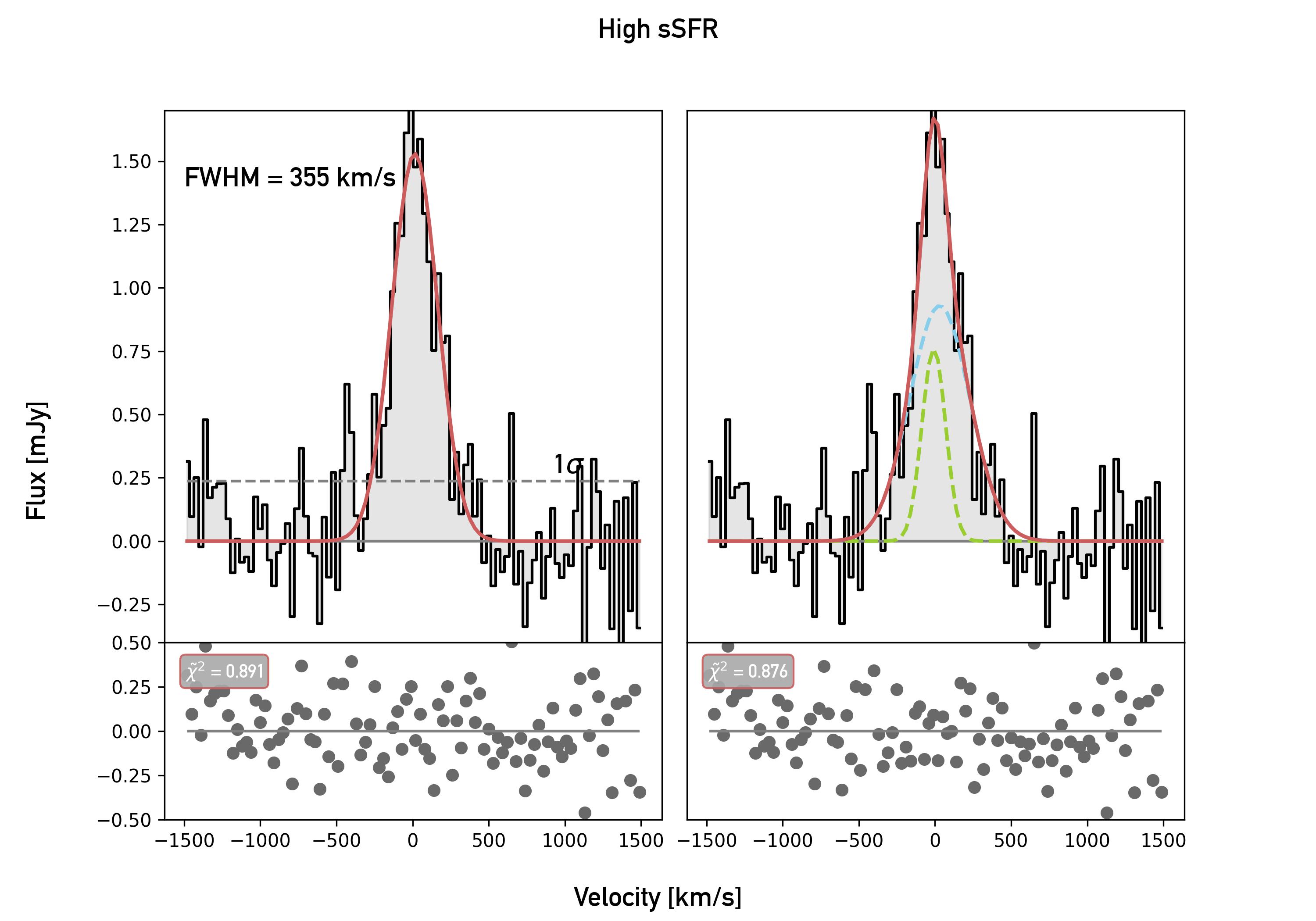}
        \label{fig:plot6}
    \end{subfigure}
    \begin{subfigure}{0.45\textwidth}
        \includegraphics[width=\linewidth]{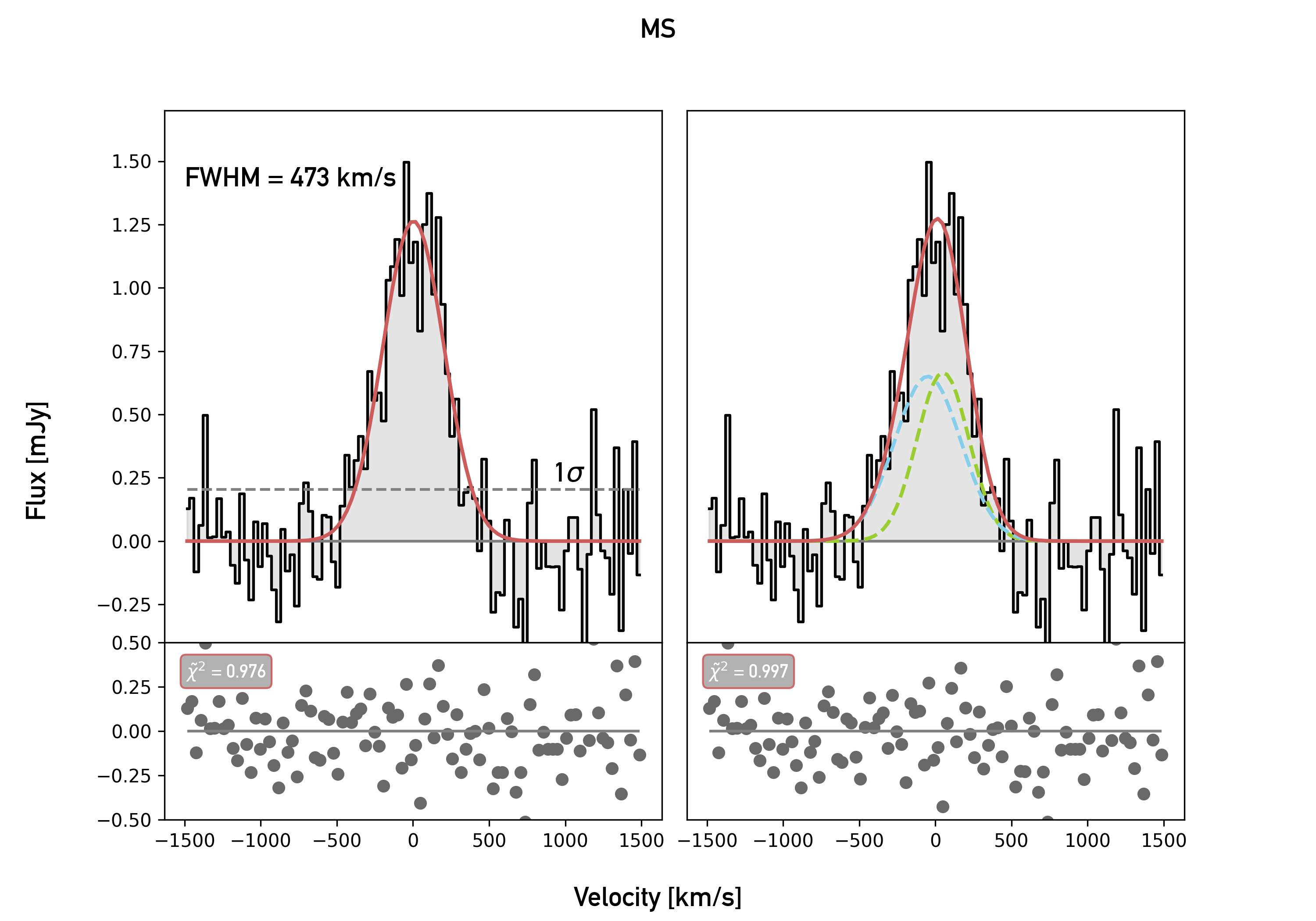}
        \label{fig:plot7}
    \end{subfigure}
    \begin{subfigure}{0.45\textwidth}
        \includegraphics[width=\columnwidth]{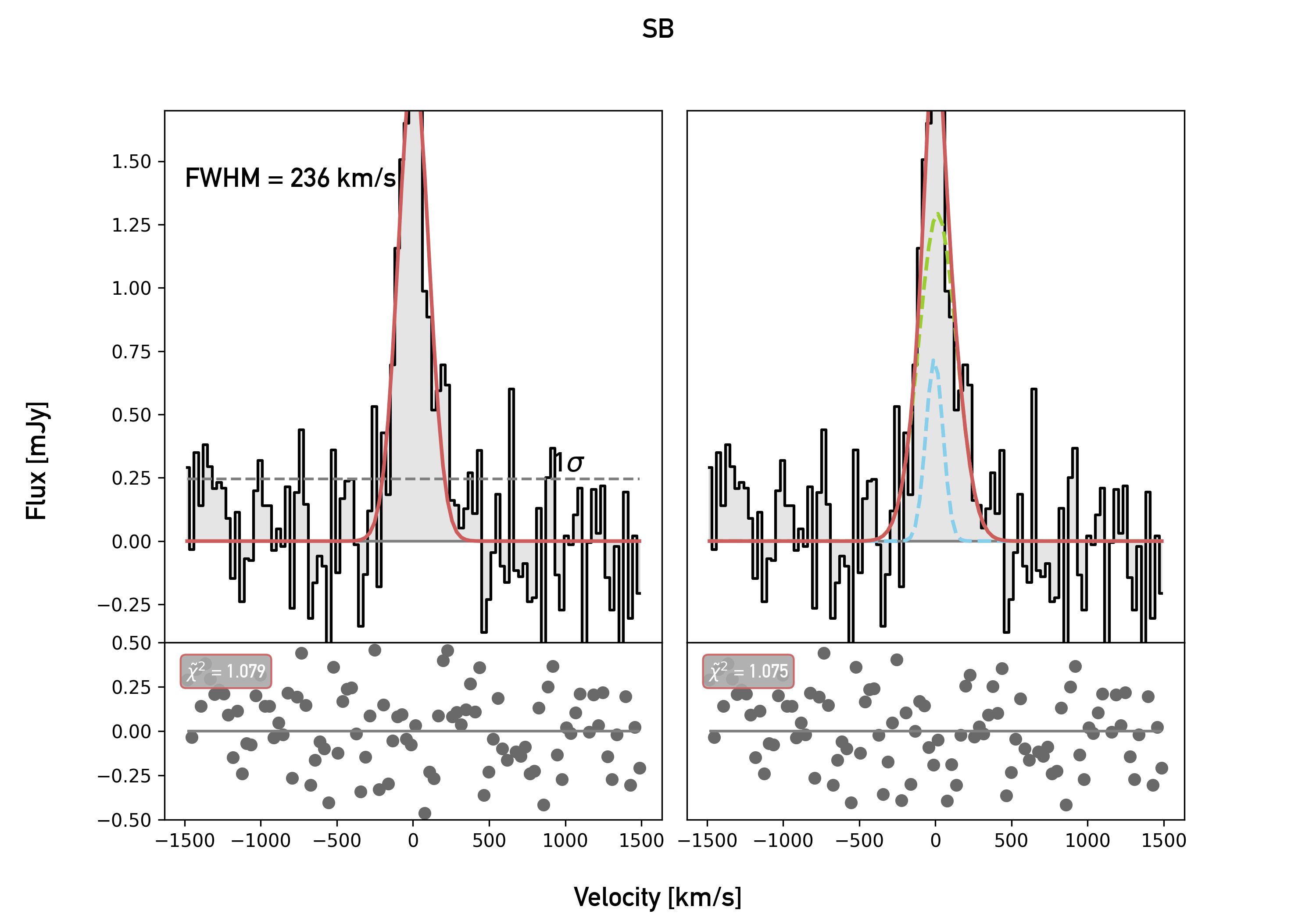}
        \label{fig:plot8}
    \end{subfigure}
    \vspace{0.5cm}
    \caption{Spectra (with natural binning of 30 km$^{-1}$) stacks in galaxy property bins. From top to bottom, we show the stellar mass bin, the infrared luminosity bin, the specific star formation rate bin, and the distance to the MS bin. The first panel of the left and right columns shows single-Gaussian fits, while the second panel shows two-Gaussian fits.}
    \label{fig:plots}
\end{figure*}

\end{document}